\documentclass[iop,numberedappendix]{emulateapj}

\usepackage{amsmath}
\usepackage{psfig}
\usepackage{apjfonts}
\usepackage{color}

\def\jcap{JCAP}
\def\icarus{Icarus}

\def\cm{\textrm{cm}}
\def\km{\textrm{km}}
\def\gram{\textrm{g}}
\def\dyn{\textrm{dyn}}
\def\erg{\textrm{erg}}
\def\kpc{\textrm{kpc}}
\def\pc{\textrm{pc}}
\def\Mpc{\textrm{Mpc}}
\def\Gpc{\textrm{Gpc}}
\def\eV{\textrm{eV}}
\def\keV{\textrm{keV}}

\def\GeV{\textrm{GeV}}

\def\PeV{\textrm{PeV}}
\def\EeV{\textrm{EeV}}
\def\ZeV{\textrm{ZeV}}
\def\YeV{\textrm{YeV}}
\def\XeV{\textrm{XeV}}
\def\WeV{\textrm{WeV}}
\def\VeV{\textrm{VeV}}
\def\sec{\textrm{s}}

\def\nsec{\textrm{ns}}
\def\yr{\textrm{yr}}

\def\Myr{\textrm{Myr}}
\def\Gyr{\textrm{Gyr}}
\def\Tyr{\textrm{Tyr}}

\def\kHz{\textrm{kHz}}
\def\MHz{\textrm{MHz}}

\def\nGauss{\textrm{nG}}
\def\muGauss{\mu\textrm{G}}
\def\mGauss{\textrm{mG}}
\def\Gauss{\textrm{G}}
\def\Kelv{\textrm{K}}
\def\mbarn{\textrm{mb}}
\def\sr{\textrm{sr}}

\def\mPa{\textrm{mPa}}
\def\Pa{\textrm{Pa}}

\def\dB{\textrm{dB}}
\def\gcm2{\textrm{g\ cm}^{-2}}
\def\rhoUnits{\textrm{g\ cm}^{-3}}
\def\dyncm2{\textrm{dyn\ cm}^{-2}}
\def\cms2{\textrm{cm\ s}^{-2}}
\def\atm{\textrm{bar}}
\def\Msun{\textrm{M}_{\odot}}
\def\Lsun{\textrm{L}_{\odot}}
\def\Rsun{\textrm{R}_{\odot}}
\def\Earthyr{(\oplus\ \textrm{yr})}
\def\ga{\gtrsim}
\def\la{\lesssim}
\newcommand{\mean}[1]{\ensuremath{\langle #1 \rangle}}

\shorttitle{YeV SETI}
\shortauthors{Lacki}

\begin{document}

\title{SETI at Planck Energy: When Particle Physicists Become Cosmic Engineers}
\author{Brian C. Lacki}
\affil{Institute for Advanced Study, Einstein Drive, Princeton, NJ 08540, USA; brianlacki@ias.edu}

\begin{abstract}
What is the meaning of the Fermi Paradox -- are we alone or is starfaring rare?  Can general relativity be united with quantum mechanics?  The searches for answers to these questions could intersect.  It is known that an accelerator capable of energizing particles to the Planck scale requires cosmic proportions.  The energy required to run a Planck accelerator is also cosmic, of order $100\ \Msun c^2$ for a hadron collider, because the natural cross section for Planck physics is so tiny.  If aliens are interested in fundamental physics, they could resort to cosmic engineering for their experiments.  These colliders are detectable through the vast amount of ``pollution'' they produce, motivating a YeV SETI program.  I investigate what kinds of radiation they would emit in a fireball scenario, and the feasibility of detecting YeV radiation at Earth, particularly YeV neutrinos.  Although current limits on YeV neutrinos are weak, Kardashev 3 YeV neutrino sources appear to be at least 30--100 Mpc apart on average, if they are long-lived and emit isotropically.  I consider the feasibility of much larger YeV neutrino detectors, including an acoustic detection experiment that spans all of Earth's oceans, and instrumenting the entire Kuiper Belt.  Any detection of YeV neutrinos implies an extraordinary phenomenon at work, whether artificial and natural.  Searches for YeV neutrinos from any source are naturally commensal, so a YeV neutrino SETI program has value beyond SETI itself, particularly in limiting topological defects.  I note that the Universe is very faint in all kinds of nonthermal radiation, indicating that cosmic engineering is extremely rare.
\end{abstract}


\section{Introduction}

\emph{We are here.  They are not.}  These appear to be the basic observations that we have so far in the Search for Extraterrestrial Intelligence (SETI).  On the one hand, we evolved on Earth, demonstrating that beings with the technology to communicate with other stars arise with nonzero probability.  The vastness of the Universe in space and time then suggests that extraterrestrials (ETs) evolved somewhere else.  Reasonable guesses with the Drake equation imply that it could have happened millions of times in our Galaxy alone \citep{Sagan63,Tarter01}.

On the other, we face the Fermi Paradox or the ``Great Silence'' \citep{Hart75,Brin83,Cirkovic09}.  There seems to be no airtight reason why starfaring societies cannot reach Earth or build structures that could be observed from Earth, and they have had billions of years to do it.  Although our current spacecraft are far too slow, a nuclear rocket can reach nearby stars within about a century of flight time \citep{Crawford90}.  Supposing that we could establish permanent, self-sufficient outposts around the closest stars, ones with their own industrial bases, those abodes could then send their own starships to more distant stars, and so on.  As the number of stars settled grows, the wave of starfarers expands at a few percent of $c$.  With these assumptions, we would pervade the Galaxy within $\sim 1$--$100\ \Myr$, less than a percent of the Galaxy's age.  If extraterrestrials starfarers arose within the $\sim$12 Gyr history of the Galaxy, they could have done the same, so that they would have some presence in the Solar System itself \citep{Hart75,Wright14-SF}.  

Just as humanity is transforming the Earth's geology, ecology, and climate \citep{Crutzen02}, it has been hypothesized that starfarers could alter entire astrophysical environments \citep{Kardashev85}.  Cosmic engineering is the deliberate alteration of astrophysical bodies.  It could take many forms: planet-sized geometrical constructions orbiting stars \citep{Arnold05} or pulsars \citep{Chennamangalam15}, Solar System scale asteroid mining \citep{Forgan11}, altering the movement of stars \citep{Badescu06}, structures surrounding stars to capture their entire luminosity \citep{Dyson60}, or altering stellar structure itself \citep{Criswell85,Learned08}.  In the most extreme possibilities, cosmology itself could be disturbed \citep{Olson14}.  Some of these proposals are motivated by \emph{communication}, in which aliens wish to contact others by building easily observed structures \citep{Kardashev64,Learned08,Chennamangalam15}.  Others are motivated by \emph{consumption}, in which aliens wish to maximize their population, available energy, or computational power.  But whatever the underlying motivation, cosmic engineering is relatively easy to see, across intergalactic distances for the largest projects \citep{Kardashev85}.  Nothing seems to forbid cosmic engineering, and ETs have had ample time to alter the Galaxy around us.  They apparently should not only be here, but we practically wouldn't even have to look for them.

Yet there are no credible and unambigious signs of ETs \emph{anywhere} so far, whether in the Solar System, elsewhere in the Galaxy, or in other galaxies.  This is despite a number of surveys being carried out with a variety of methods.  Radio searches are the primary tool of SETI, particularly focusing on narrowband signals \citep{Cocconi59,Tarter01}.  Although these turned up a few intriguing candidate artificial signals, none of them have ever been detected again \citep{Gray94,Gray01,Lazio02,Gray02}, and overall no compelling candidate ETs have been found \citep{Blair92,Horowitz93,Shostak96,Rampadarath12}.  Optical and near-infrared laser pulses are a plausible method of interstellar communication, even with our current technology \citep{Schwartz61}, and optical SETI programs aim to detect them \citep{Eichler01,Howard04,Hanna09}.  There have been a few attempts to search for artifacts in the Solar System as well, to check whether they \emph{are} in fact here \citep{Freitas80,Freitas83,Steel95}.  Finally, cosmic engineering is thought to be detectable through its waste heat \citep{Dyson60,Kardashev64,Sagan66}, so far- to mid-infrared surveys constrain its prevalence \citep{Slysh85,Carrigan09,Wright14-Results}.  

Other channels for SETI have been proposed, including: looking for artificial planetary transits \citep{Arnold05}, deliberate stellar variability \citep{Learned08}, or industry \citep{Whitmire80,Loeb12,Lin14}; more thorough searches for artifacts on planetary surfaces, interplanetary space, or starships \citep{Bracewell60,Rose04,Harris02,Davies12}; and searching for messages sent as high energy neutrinos \citep{Subotowicz79,Learned94,Silagadze08,Learned09} or cosmic rays (CRs; \citealt{Swain06}).  Neutrino communication in particular has been demonstrated on Earth \citep{Stancil12}.  Nothing artificial has stood out yet with these methods either. 

So, are we alone in the Galaxy, or even the observable Universe?  Or does something always prevent ETs from starfaring and cosmic engineering -- with the ominous implication that humanity has no chance of achieving starfaring either?  

The Fermi Paradox, with its pessimistic implications for SETI and its frightening insinuations about us, provokes a lot of controversy \citep{Cirkovic09}.  The first issue is whether interstellar flight and settlement is feasible.  But the ETs may not even have to physically send astronauts to other stars.  As the notorious counterargument goes, they might launch replicating machines in their stead -- yet we do not see signs of these machines either \citep{Tipler80}.  The other main issue is whether ETs decide to settle other stars and begin cosmic engineering.  As often phrased, the Fermi Paradox seems to assume that ETs are interested in consuming the Galaxy for the mere sake of consumption.  Might ETs simply have different values that discourage this kind of expansion \citep[c.f.,][]{Anderson60,Sagan83,HaqqMisra09}?  

The problem is that everyone must agree not to go starfaring.  Even a single species of ETs may be composed of a multitude of societies, themselves composed of a multitude of groups, with varying goals over many historical epochs \citep{Hart75,Wright14-SF}.  After all, our own societies include a wide range of opinions on whether ETs exist, how we might find them, and particularly whether we should contact them.  Arguments from large numbers are turned on their head; if it defies belief that we are alone among millions of stars, it defies belief that there are no cosmic engineers or starfarers among millions of societies.  The Fermi Paradox highlights an instability; as with a hydrodynamic instability, only one unstable mode can bring everything crashing down.  But is there a good reason to cosmic engineer, something to trigger the instability? 

I propose \emph{curiosity} as a motivation for cosmic engineering.  The other great quest in physics and astronomy of the past few decades has been the search for a unified theory of physics including gravity.  The most natural energy scale for unification is the Planck energy,\footnote{The SI prefixes for very large numbers are E ($10^{18}$; \emph{exa}), Z ($10^{21}$; \emph{zetta}), and Y ($10^{24}$; \emph{yotta}).  No larger SI prefixes have been decreed, but I adopt symbols that continue in reverse alphabetical order: X ($10^{27}$), W ($10^{30}$), and V ($10^{33}$), as proposed (sometimes in jest) by Jeff Aronson, Jim Blower, and Sbiis Saiban.  The history of these systems is described by \citet{Saiban15}.} $E_{\rm Planck} = \sqrt{\hbar c^5 / G}$, which evaluates to $1.22 \times 10^{28}\ \eV = 12.2\ \XeV$ -- far, far beyond the reach of any accelerator possible with our technology.  While various extensions to the Standard Model of particle physics have been proposed, there seems to be no evidence of any of them in the Large Hadron Collider (LHC; \citealt{Chatrchyan12,Aad13}) or in other precision tests (many searches are listed in \citealt{Olive14}).  This leads to speculations that we are living in the ``Nightmare Scenario'', in which there is no new physics until the Planck scale itself \citep{Cho07}.  Of course, we know that there is physics beyond the Standard Model in the form of dark matter and dark energy.  But we have no guarantee that they can be solved without Planck-scale physics.  Or -- maybe just as bad -- they might turn out to be just another set of fields, providing no revolutionary insights on how to unify fundamental physics.  We appear to be facing a physics ``desert'' between 1 TeV and 10 XeV, with no observations telling us how to get across it.

In discussions of the Nightmare Scenario and the physics desert, the idea of probing Planck scale physics directly is generally dismissed because it would require a particle accelerator that is ``as big as the Solar System'', ``the size of the Galaxy'', or ``the size of the whole Universe'' \citep{Akahito89,Greene99,Davies03,Susskind08,Kaku10,Adler10}.  But starfaring ETs conceivably could build an accelerator the size of a galaxy over a few million years.  If ETs are interested in fundamental physics, they may resort to such engineering.  So why not look for artificial particle accelerators that are \emph{literally} the size of a galaxy?

Aside from resolving the accelerators themselves, or seeing their waste heat, we could search for YeV--XeV radiation generated within the accelerator.  Neutrinos could be particularly effective in escaping the accelerator and reaching the Earth.\footnote{When I refer to ``neutrinos'' in this paper, I actually mean both neutrinos ($\nu$) and anti-neutrinos ($\bar{\nu}$), and I include all three known flavors (electron, muon, and tau).  For YeV--XeV neutrino experiments, the differences between these types do not matter much.}  The natural background of ultra-high energy cosmic rays (UHECRs) falls off above $\sim 40\ \EeV$, and none have been observed with an energy greater than $\sim 1\ \ZeV$ \citep{Abbasi08,Abraham08,Abraham10}.  Although YeV--XeV cosmic rays (CRs) might be emitted by Planck-scale relics of the early Universe like cosmic strings \citep[e.g.,][]{Hill87,Birkel98,Berezinsky09}, these ``top down'' models are strongly constrained \citep{Rubtsov06}.  Detection of YeV--XeV radiation would be a revolutionary discovery, implying either a completely new class of accelerators or new physics \citep{Thompson11}.

\begin{deluxetable}{lll}
\tablecaption{Constants in Gaussian centimeter-gram-second units}
\tablehead{\colhead{Name} & \colhead{Value} & \colhead{Explanation}}
\startdata
$c$               & $2.998 \times 10^{10}\ \cm\ \sec^{-1}$             & Speed of light in vacuum\\
$G$               & $6.674 \times 10^{-8}\ \dyn\ \cm^2\ \gram^{-2}$    & Newton's gravitational constant\\
$h$               & $6.626 \times 10^{-27}\ \erg\ \sec$                & Planck's constant\\
$\hbar$           & $1.055 \times 10^{-27}\ \erg\ \sec$                & $h / (2 \pi)$\\
$m_e$             & $9.109 \times 10^{-28}\ \gram$                     & Rest mass of an electron\\
$m_p$             & $1.673 \times 10^{-24}\ \gram$                     & Rest mass of a proton\\
$e$               & $4.80 \times 10^{-10}\ \erg^{1/2}\ \cm^{1/2}$      & Electric charge of an electron\\
$E_{\rm Planck}$      & $1.957 \times 10^{16}\ \erg$         & Planck energy ($\sqrt{\hbar c^5 / G}$)\\
$\sigma_{\rm Planck}$ & $3.284 \times 10^{-65}\ \cm^2$    & Planck cross section ($2 G h / c^3$)\\
$L_E$             & $3.629 \times 10^{59}\ \erg\ \sec^{-1}$             & Einstein luminosity ($c^5/G$)\\
$\alpha_{\rm EM}$ & $0.007297$                                         & Fine-structure constant ($e^2/(\hbar c)$)\\
$a_{\rm SB}$      & $7.573 \times 10^{-15}\ \erg\ \cm^{-3}\ \Kelv^{-4}$ & Rescaled Stefan-Boltzmann constant\\
$\Rsun$           & $6.961 \times 10^{10}\ \cm$                         & Radius of the Sun\\
$\Msun$           & $1.988 \times 10^{33}\ \gram$                      & Mass of the Sun\\
$\Lsun$           & $3.846 \times 10^{33}\ \erg\ \sec^{-1}$            & Solar luminosity (electromagnetic)\\
$\yr$             & $3.1557 \times 10^7\ \sec$                         & Earth year\\
$\atm$            & $10^6\ \dyn\ \cm^{-2}$                             & Approximate air pressure at sea level\\
$\eV$             & $1.602 \times 10^{-12}\ \erg$                      & Electron-volt\\
barn              & $10^{-24}\ \cm^2$                                  & Barn (cross section unit)
\enddata
\tablecomments{I use the physical constants given in \citet{Olive14}.}
\label{table:cgsConstants}
\end{deluxetable}

Because the paper is about individual relativistic particles, I use Gaussian centimeter-gram-second units (\citealt{Jackson98}, p. 777).  The symbols for the constants of nature appear in Table~\ref{table:cgsConstants}.  I also rate ET artifacts' power output $L$ with the Kardashev scale as quantified by \citet{Sagan73}:
\begin{equation}
K = 2.0 + 0.1 \log_{10} \left(\frac{L}{\Lsun}\right).
\end{equation}
So a power of $1\ \Lsun$ is given as Kardashev 2.0 (K2.0 or K2), a power of $10^5\ \Lsun$ is Kardashev 2.5 (K2.5), a power of $10^{10}\ \Lsun$ is Kardashev 3.0 (K3.0 or K3), and so on.\footnote{For $K$ different than 2, these values are different than those used by \citet{Kardashev64}.}

\section{Are Planck accelerators even possible?}

There are two broad ways to generate Planck energy particles.  In a ``top-down'' approach, they are made when a Planck scale object, such as a heavy particle or a topological defect (like a magnetic monopole or cosmic string) decays \citep[e.g.,][]{Hill87,Birkel98,Berezinsky09}.  We do not know whether these actually exist, though.  The approach we use in our colliders is ``bottom-up'', in which light particles are energized all the way up to Planck scale.

Of the four known fundamental forces, we use only electromagnetism to accelerate particles.  The ability to accelerate a particle partly depends on the force coupling constant $\alpha$, which is $\alpha_{\rm EM} = 0.007$ for electromagnetism.  The weak nuclear force has a slightly higher coupling constant, and the strong nuclear force has $\alpha \ga 1$, but their ranges are far too short to use in macroscopic accelerators.  Gravitation has negligible strength for the fundamental particles, with a coupling constant of $\alpha_g = G m^2 / (\hbar c)$ for particles of mass $m$ (about $10^{-38}$ for protons; \citealt{Gould85,Rothman06}), and is useless in our current accelerators.  General relativistic effects may allow particle acceleration to Planck energy and beyond very close to black holes, but whether this ever actually happens is debated \citep{Piran75,Banados09,Berti09,Jacobson10,Wei10}.

A fundamental problem with trying to build electromagnetic Planck accelerators is that they tend to collapse into black holes, because the electromagnetic energy density within them is so high \citep{Thompson11}.  Assuming the magnetic fields fill a spherical region, the limiting particle energy is $\sqrt{\alpha_{\rm EM}} E_{\rm Planck} \approx 1\ \XeV$ \citep{Kardashev95,Thompson11}.  There are factors of order unity an ET might exploit to push this limit.  For example, if the magnetic field confining particles in the collider fills a thin tube instead of a sphere, perhaps Planck energies could be achieved (although these configurations could be unstable).  If ETs used fundamental particles with greater charge, then this limit is also relieved.  On the other hand, energy losses from synchrotron and curvature emission sets an upper limit of just 0.1 YeV \citep{Thompson11}.  The magnetic fields in actual astrophysical objects limit nucleon energies to $\sim 1\ \ZeV$, as the magnetic fields are far too small to create a black hole \citep{Hillas84}.  

\citet{Casher95,Casher97} surmised that the laws of physics forbid us from seeing a Planck particle, arguing no accelerator can be built that reaches those energies.  A necessary assumption is that macroscopic amounts of matter cannot reach ultrarelativistic speeds.  Otherwise, a particle that ``merely'' reaches ZeV to YeV energies in some ultrarelativistic flow can reach Planck energy in our frame (or vice-versa).  But since then, we have strong evidence that gamma-ray bursts (GRBs) launch flows with bulk Lorentz factors $\Gamma \approx 100$--$1000$ \citep{Lithwick01,Abdo09-GRB}.  There appears to be no fundamental law against accelerating YeV particles, which are boosted to Planck energies in the flow's frame.  As long as Lorentz invariance holds to Planck scale, reaching Planck energy is technically allowed if extremely difficult.  Relativistic boosting also is a possible way around the bounds in \citet{Kardashev95} and \citet{Thompson11}, if the entire central engine itself is boosted to large Lorentz factors.  (This is how cosmic strings might produce Planck energy particles, for example; \citealt{Berezinsky09}.)

A subtlety for colliders built to investigate Planck scale physics is that the relevant quantity is the Mandelstam $s$ (center of mass energy squared), not just the particle energy $E$ (chapter 46 of \citealt{Olive14}).  Because of special relativity, if a relativistic Planck energy particle hits a target particle with mass $m$ at rest, $\sqrt{s}$ is only about $\sqrt{E_{\rm Planck} m c^2}$.  Two particles with Planck scale energy must hit each other in the lab frame to reach $\sqrt{s} \approx E_{\rm Planck}$.  This may be done by accelerating two beams of particles and aiming them at each other.  

In any case, it appears that (bottom-up) Planck accelerators require a lot of fine-tuning, and are unlikely to appear in nature although ETs might find ways to engineer them.  Of course, even if Planck energy is unachievable, ETs could still be interested in physics at smaller YeV grand unified scales.  The basic arguments of this paper still apply, though with weaker constraints on the size and luminosity of accelerators.

\section{Why they should be big}
\label{sec:BigAccelerators}

Electric fields ${\cal E}$ accelerate charged particles to higher energies, whereas magnetic fields $B$ merely deflect them without doing work.  The dot product of ${\cal E}$ and $B$ is Lorentz invariant; the component of ${\cal E}$ parallel to the magnetic field is the same in all frames, but there is some frame where there is no electric field perpendicular to the magnetic field.  If $|{\cal E} \cdot B| > 0$, then the accelerator is basically electrostatic -- the particle accelerates along the electric field, gaining energy as it does so.  Otherwise, the scattering of the particle by the electromagnetic field is completely elastic in some frame, with no work done on it; the relativistic boost of $B$ is fundamentally the reason for energy transfer to the particle.

The theory of quantum electrodynamics (QED) defines a characteristic electromagnetic field ${\cal B}_{\rm QED}^e = m_e^2 c^3 / (e \hbar) = 4.4 \times 10^{13}\ \Gauss$ \citep{Harding06}.  The characteristic size of a particle accelerator with electromagnetic fields ${\cal B}$  that can accelerate a particle of charge $Ze$ to energy $E$ is 
\begin{equation}
\label{eqn:AcceleratorSize}
%E_Pl = sqrt(h c / (2 pi G)) c^2 = 1.22e28 eV
l_{\rm min} = \frac{E}{Z e {\cal B}_{\rm QED}^e} \approx 9.3 \times 10^{11}\ \cm\ \left(\frac{E}{E_{\rm Planck}}\right) \left(\frac{\cal B}{{\cal B}_{\rm QED}}\right)^{-1}.
\end{equation}
The characteristic size of a Planck-scale accelerator is greater than $10\ \Rsun$.

For an electrostatic accelerator (like a linear accelerator), this is truly a fundamental limit as far as we know.  The QED vacuum becomes unstable if ${\cal E} > {\cal B}_{\rm QED}^e$ unless there is a suitable magnetic field.  Any attempt to build the electric field to ${\cal B}_{\rm QED}$ would simply result in the creation of electron-positron pairs instead \citep{Fedotov10}.   An ET that wished to build an electrostatic accelerator would necessarily be spacefaring because the accelerator is too big to fit onto any planet.

With a magnetic accelerator (like a synchrotron), equation~\ref{eqn:AcceleratorSize} corresponds to the Hillas criterion \citep{Hillas84}.  This says that the particle's gyroperiod must be smaller than the light-crossing time of the accelerator region.  In principle, the accelerator could be smaller than $\sim 10\ \Rsun$ if $B \gg {\cal B}_{\rm QED}^e$, a condition reached in magnetars. 

No Planck accelerator made of normal atomic matter could be as small as equation~\ref{eqn:AcceleratorSize}.  Atoms deform in magnetic fields of $\sim 10^9\ \Gauss$ \citep{Harding06}.  A Planck accelerator with gigagauss electromagnetic fields would be several thousand AUs wide.  A minimum electromagnetic field of $\sim 4\ \mGauss$ within the accelerator is set by the size of the Universe, $\sim 3\ \Gpc \approx 10^{28}\ \cm$.  This kind of magnetic field is far bigger than those within the intergalactic medium (IGM; \citealt{Dolag05,Neronov10,Yoon14}) or within most galaxies \citep[e.g.,][]{Lacki10-FRC1,Beck12}.   Based on the required electromagnetic energy density, a Planck accelerator is probably close to collapsing into a black hole, and black holes might be harnessed \citep[c.f.,][]{Kardashev95,Vidal11,Thompson11}.

\section{Why they should be bright}
\label{sec:BrightAccelerators}
Doing experiments with particle accelerators is not just a matter of reaching the highest energies possible.  The other central consideration is the particle luminosity of the accelerator, essentially the number flux of particle bunches as they pass through each other.\footnote{I will consistently refer to this as ``particle luminosity''.  ``Luminosity'' on its own refers to the astrophysical meaning of energy luminosity in this work.} Accelerators are built to find events, particle interactions regulated by physics.  Each type of event has a cross section $\sigma$.  The rate at which a class of events occur in the accelerator is proportional to the product of the cross section and the particle luminosity.  Thus, the integrated particle luminosity over time is basically the reciprocal of a cross section; the longer the accelerator runs, the more the effective cross sections probed decrease.

The ``natural'' cross section for quantum processes occurring at an energy $E$ is
\begin{equation}
\sigma_{\rm natural} \sim 4 \pi \left(\prod \alpha^n\right) \left(\frac{\hbar c}{E}\right)^2 = \frac{1}{\pi} \left(\prod \alpha^n\right) \left(\frac{h c}{E}\right)^2
\end{equation}
where the $\alpha$ are coupling constants of each force involved, and $n$ is the order of each coupling, the number of each kind of vertex in the Feynman diagram \citep[e.g.,][]{Halzen84}.\footnote{Extrapolating to Planck energy, $\alpha_g = 1$.}  When $E$ is the mass energy of some force carrier, then $(hc/E)$ is its Compton wavelength.  At Planck energy and $\alpha \approx 1$, the natural cross section is approximately the Planck area:
\begin{equation}
\sigma_{\rm Planck} \equiv 2 \frac{G h}{c^3} = 3.3 \times 10^{-65}\ \cm^{2} 
\end{equation}
Note that this is approximately the surface area of a black hole with Planck mass, for which quantum gravity effects should be strong \citep[c.f.,][]{Dimopoulos01}.  It is also the approximate cross section for first-order gravitational scattering occurring at low energy \citep[e.g.,][]{Papini77,Saif91,Rothman06}.  Attaining the Planck cross section requires an incredible integrated particle luminosity, about $10^{24}$ times greater than that of the LHC so far.  On its own, this requires either a huge number of particles being accelerated, a large number of crossings between particle bunches, or extremely dense bunches.

Unfortunately for would-be Planck accelerator builders, there are always other interactions with much greater cross sections that drain the particles of their energy.  These serve as a foreground for the sought Planck-physics events; the detectors necessarily sift through many events to detect the few that are interesting.  As such, accelerators built to detect Planck events are extremely wasteful and produce vast amounts of ``pollution''.  The minimum amount of energy needed to run the accelerator is
\begin{equation}
\label{eqn:NeededEnergies}
\Upsilon \ga n_{\rm events} \kappa E \frac{\sigma_{\rm any}}{\sigma_{\rm natural}},
\end{equation}
where $n_{\rm events} \ga 1$ is the number of Planck-scale events sought, $\kappa$ is the inelasticity of a typical collision, and $\sigma_{\rm any}$ is the cross section for any kind of interaction between the beam particles.

Hadron-hadron colliders like the LHC are the easiest to build because they avoid synchrotron energy losses, but they are especially wasteful.  In our current understanding of QCD, the typical cross section of $pp$ collisions is $3 \times 10^{-26}\ \cm^{-2}$ and increases with energy to about $10^{-24}\ \cm^{-2}$ at Planck scale, with an inelasticity of $\sim 0.1$ to 0.2 \citep{Olive14}.  Only 1 in $10^{41}$ collisions would contain the signatures of Planck-scale physics, and each of those collisions involves a proton that has been accelerated to Planck energy.  The total amount of energy is $\sim 1 \times 10^{56} (E/E_{\rm Planck})^3 \ \erg$, a hundred times the entire mass-energy of the Sun.  Leptonic colliders with electrons or muons as beam particles are limited by pair production processes that also have cross sections of a barn.  Thus, Planck-scale accelerators require cosmic engineering, simply to ensure that there are enough events that are interesting.  

The situation is actually worse than this for hadronic colliders.  Hadrons contain partons (quarks, antiquarks, and gluons) that carry only a fraction of the energy of the entire hadron, and it is the partons that actually undergo reactions.  Almost all of the collisions will be between the vast multitudes of ``sea'' partons with a negligible center-of-mass momentum (with the $x$ variable close to $0$), instead of the three valence quarks (with $x \approx 1$; \citealt{Halzen84}).  So the parton-level collision energy $\sqrt{\hat{s}}$ in most hadronic collisions is much smaller than the hadron-level $\sqrt{s}$ (\citealt{Quigg11}; see also the parton particle luminosity functions in Figures 32 -- 56 of \citealt{Eichten84} and Figure 78 of \citealt{Campbell07}).  Planck-scale physics in interactions between partons ($\sqrt{\hat{s}} = E_{\rm Planck}$) would be very rare indeed.  

\begin{deluxetable*}{lcccccccccc}
\tablecaption{Accelerator power to reach $\sigma_{\rm Planck}$}
\tablehead{\colhead{Beams} & \multicolumn{3}{c}{Limiting interaction\tablenotemark{a}} & \multicolumn{2}{c}{$\Upsilon$ \tablenotemark{b}} & \multicolumn{2}{c}{$\Upsilon / t_H$} & \multicolumn{2}{c}{$\Upsilon / \Myr$} & \colhead{Notes} \\ & \colhead{Process} & \colhead{$\sigma (s_{\rm Planck}) [{\mbarn}]$} & \colhead{${\cal O}[\sigma(s)]$} & \colhead{$(\erg)$} & \colhead{$(\Msun c^2)$} & \colhead{$(\erg\ \sec^{-1})$} & \colhead{$(\Lsun)$} & \colhead{$(\erg\ \sec^{-1})$} & \colhead{$(\Lsun)$} & }
\startdata
%All scaled to kappa = 0.1
$pp$               & $p + p \to \pi + {\rm anything}$                  & $\sim 2000$          & $\ln^2 s$ (?) & $1 \times 10^{56}$ & 70                 & $3 \times 10^{38}$ & $7 \times 10^4$    & $4 \times 10^{42}$ & $1 \times 10^9$ & (c)\\
$p\gamma$          & $p + \gamma \to \pi + {\rm anything}$             & $\sim 7$             & $\ln^2 s$ (?) & $4 \times 10^{53}$ & 0.2                & $9 \times 10^{35}$ & 200                & $1 \times 10^{40}$ & $3 \times 10^6$ & (d)\\
$e^{\pm}e^-$       & $e^{\pm} + e^- \to e^+ + e^- + e^+ + e^-$         & $1500$               & $\ln^3 s$     & $9 \times 10^{55}$ & 50                 & $2 \times 10^{38}$ & $5 \times 10^4$    & $3 \times 10^{42}$ & $7 \times 10^8$ & (e)\\
$\mu^{\pm}\mu^-$   & $\mu^{\pm} + \mu^- \to \mu^+ + \mu^- + e^+ + e^-$ & $1000$               & $\ln^3 s$     & $6 \times 10^{55}$ & 30                 & $1 \times 10^{38}$ & $3 \times 10^4$    & $2 \times 10^{42}$ & $5 \times 10^8$ & (e)\\
$\gamma\gamma$     & $\gamma + \gamma \to e^+ + e^- + e^+ + e^-$       & $0.00645$            & $1$           & $4 \times 10^{50}$ & $2 \times 10^{-4}$ & $9 \times 10^{32}$ & 0.2                & $1 \times 10^{37}$ & 3000            & (g)\\
                   & $\gamma + \gamma \to {\rm hadronic}$              & $\sim 0.02$          & $\ln^2 s$ (?) & $1 \times 10^{51}$ & $7 \times 10^{-4}$ & $3 \times 10^{33}$ & 0.7                & $4 \times 10^{37}$ & $1 \times 10^4$ & (d)\\
$p\nu$             & $p + \nu \to {\rm anything}$                      & $0.059$          & $\sqrt{s}^{0.36}$ & $3 \times 10^{51}$ & 0.002              & $8 \times 10^{33}$ & 2                  & $1 \times 10^{38}$ & $3 \times 10^4$ & (h)\\
$\nu_i\bar{\nu}_j$ & $\nu_i + \bar{\nu}_j \to \ell_i + \bar{\ell}_j$   & $2.2 \times 10^{-7}$ & $1$           & $1 \times 10^{46}$ & $7 \times 10^{-9}$ & $3 \times 10^{28}$ & $8 \times 10^{-6}$ & $4 \times 10^{32}$ & 0.1             & (i)\\
$\nu_i \nu_j$      & $\nu_i + \nu_j \to \nu_i + \nu_j$                 & $7.0 \times 10^{-8}$ & $1$           & $4 \times 10^{45}$ & $2 \times 10^{-9}$ & $9 \times 10^{27}$ & $2 \times 10^{-6}$ & $1 \times 10^{32}$ & 0.03            & (i)
\enddata
\label{table:AcceleratorPower}
\tablenotetext{a}{The cross sections are evaluated at $s = E_{\rm Planck}^2$.}
\tablenotetext{b}{The total energies needed are calculated from eqn.~\ref{eqn:NeededEnergies}, assuming that $n_{\rm events} = 1$, $\kappa = 0.1$, and $\sigma_{\rm natural} = \sigma_{\rm Planck}$.}
\tablenotetext{c}{Cross section as estimated from \citet{Olive14} (Section 50); the inelastic cross section is of order half of this.  The $\ln^2 s$ dependence is expected in ``black disk'' models of nucleons.  The Froissart bound is frequently interpreted as leading to a $\ln^2 s$ dependence as $s \to \infty$.  But the size of the $\ln^2 s$ term, as well as the possibility of a slightly more rapidly growing term, are disputed.  See, for example, \citet{Azimov11,Block11,Fagundes13,Anisovich13-BlackDisk,Anisovich13-Asymptotics}.}
\tablenotetext{d}{Uses the \citet{Olive14} fits to the hadronic interaction cross sections.}
\tablenotetext{e}{Formulae for cross sections summarized in \citet{Budnev75}.  See also the discussion on muon accelerators in \citet{Ginzburg96}.}
\tablenotetext{g}{From \citet{Brown73-DPP}.}
\tablenotetext{h}{From \citet{Gandhi98}.}
\tablenotetext{i}{From \citet{Roulet93}.}
\end{deluxetable*}

The minimum amounts of energy for various kinds of accelerators are given in Table~\ref{table:AcceleratorPower}, as are the reactions that should waste the most energy at Planck energies.  Not surprisingly, a neutrino-neutrino collider makes the least pollution, as the cross section for neutrino-neutrino collisions is small \citep{Gandhi98,Alikhanov08}.  Even then, the amount of energy that must be invested is one tenth of the mass energy of the Moon.  Note that these are the absolute minimum energies.  If ETs wish to get good statistics on Planck scale events, they may want to study thousands of events, requiring that much more energy.

The brightness of the accelerator, its waste heat, and its YeV--XeV radiation depends on how long it runs.  I assume the experiments would last $t_H = 4.4 \times 10^{17}\ \sec$ at longest.  Galaxies evolve on Gyr timescales, and the supplies of power and mass might vanish around the accelerator if the ETs took much longer.  A more natural timescale might be the ``Fermi'' timescale to cross a galaxy and establish a K3 society, about 1 Myr.  The luminosities for these timescales are also listed in Table~\ref{table:AcceleratorPower}.  A hadronic Planck accelerator would need to be K3 for a run time of 1 Myr, and a neutrino collider would still be K2 over that duration.

At the opposite extreme, the accelerator lasts at least one light crossing time, and the smallest an accelerator could be is a Schwarzschild radius.  The minimum run time is then $\sim \Upsilon G / c^5$, with the accelerator reaching the maximal Einstein luminosity of $L_E = c^5 / G = 3.63 \times 10^{59}\ \erg\ \sec^{-1}$ (K4.6; \citealt{Hartle03}).  These kinds of accelerators are brief transients lasting between a fraction of a millisecond and a few hours \citep[c.f.,][]{Thompson11}.

\section{How they should shine}

\subsection{Conditions in the accelerator region}
\label{sec:AcceleratorConditions}
Which kinds of radiation the accelerator could emit depends on the conditions in the accelerator region.  Any YeV particles generated in collisions could radiatively cool to low energies if the accelerator is filled with magnetic fields, radiation, or baryons.  If the densities are high enough, all of the energy could be thermalized.  But if the accelerator is optically thin to at least some YeV particles, it shines directly in YeV radiation and appears as an unprecedented phenomenon.

I base my estimates on a partial spherical shell model of internal shocks, which is used in estimating high energy radiation from relativistic transients like Gamma Ray Bursts (\citealt{Rees94,Waxman95,Thompson11}).  The outflow is powered by a central engine with luminosity $L$ and a variability timescale $\delta t$.  It emits an outflow that expands with bulk Lorentz factor $\Gamma$ that covers $4 \pi \Psi$ steradians (so $\Psi = 1$ is isotropic).  The outflow is unsteady, with different parts having slightly varying speeds.  The irregularities crash into one another and generate internal shocks when the outflow has expanded to a characteristic size $r = \Gamma^2 c \delta t$.  It is in these internal shocks that acceleration occurs.  

The accelerator region moves outward with the bulk flow.  I define the accelerator frame to be comoving with the bulk flow.  Quantities in the accelerator frame are marked with primes ($^\prime$), whereas quantities in the engine frame are unprimed.  The energy density of the flow in the accelerator frame is 
\begin{equation}
\label{eqn:uPrime}
u^{\prime} = \frac{L}{4 \pi \Psi \Gamma^2 c r^2} = \frac{L}{4 \pi \Psi \Gamma^6 c^3 \delta t^2}.
\end{equation}
The typical energy density in the flow is 
\begin{equation}
u^{\prime} = 1 \times 10^7\ \erg\ \cm^{-3}\ \Psi^{-1} \Gamma_3^{-6} \delta t_{\sec}^{-2} \left(\frac{L}{L_E}\right),
\end{equation}
where $\Gamma_3 = \Gamma / 1000$ and $\delta t_{\sec} = [\delta t / (1\ \sec)]$.  For an isotropic outflow, the highest proton energies are achieved in outflows with $\Gamma \approx 2000$ \citep{Thompson11}.  

Instead of being spherical, the outflow may be highly beamed with an opening angle $\la 1/\Gamma$.  The characteristic covering fraction is $\Psi_{\rm \Gamma} \equiv 1/(2\Gamma^2)$, and the energy density is
\begin{align}
u^{\prime} & = \frac{L}{2 \pi \Gamma^4 c^3 \delta t^2} = 2000\ \erg\ \cm^{-3}\ \psi_{\Gamma}^{-1} \Gamma_6^{-4} \delta t_{\sec}^{-2} \left(\frac{L}{L_E}\right),
\end{align}
after defining $\Gamma_6 = \Gamma / 10^6$ and $\psi_{\Gamma} = \Psi / \Psi_{\Gamma}$.  Planck particles can then be accelerated for extreme outflows with $\Gamma > 10^6$ \citep{Thompson11}.

\subsubsection{Escape from the accelerator or decay inside it?}
\label{sec:EscapeVsDecay}
The accelerator region is a shell with thickness of $\delta r^{\prime} = \Gamma c \delta t$.  The minimum possible dynamical time is just the light-crossing time of the shell, $t_{\rm dyn}^{\prime} = \Gamma \delta t = 1000\ \sec\ \Gamma_3 \delta t_{\sec}$.  This time is also the typical time it takes a CR to escape the accelerator and the maximum time it takes a CR to complete one Larmor orbit \citep{Thompson11}.

Particle accelerators naturally create a slew of unstable particles, including muons, mesons, high energy baryons, and gauge bosons (Tables~\ref{table:ChargedUnstable} and \ref{table:NeutralUnstable}).  If their lifetime is short enough, they decay before they can escape the accelerator.  Since the unstable particles are time dilated, their decay time in the accelerator frame is
\begin{align}
t_{\rm decay}^{\prime} & = \frac{E \tau}{\Gamma m c^2} = 1 \times 10^7\ \sec\ \frac{E_{28} \tau_{\nsec}}{\Gamma_3 m_{\GeV}}.
\end{align}
Here, $\tau$ is the rest lifetime of the particle and $E^{\prime}$ is the particle (kinetic) energy in the accelerator frame.  A particle observed to have energy $E$ in the engine frame has $E^{\prime} = E / \Gamma$.  For compactness, I use the auxiliary variables $\tau_{\nsec} = [\tau / (1\ \nsec)]$, $E_{28} = [E / (10^{28}\ \eV)]$, and $m_{\rm GeV} = [m / (1\ \GeV)]$.

\begin{deluxetable}{lcccc}
\tablecaption{Common unstable charged particles}
\tablehead{\colhead{Particle} & \colhead{Rest energy} & \colhead{Rest lifetime} & \colhead{$\gamma_{\rm Planck} c\tau$} & \colhead{Direct $\nu$/$\bar{\nu}$}\\ & \colhead{$(\GeV)$} & \colhead{$(\nsec)$} & \colhead{$(\pc)$}}
\startdata
$W^{\pm}$    & 80.4   & $3.07 \times 10^{-16}$ & $4.53 \times 10^{-16}$ & \checkmark\\
\cutinhead{Charged Leptons}
$\mu^{\pm}$  & 0.1057 & 2197.0                 & $2.46 \times 10^6$     & \checkmark\\
$\tau^{\pm}$ & 1.777  & $2.91 \times 10^{-4}$  & 0.0194                 & \checkmark\\
\cutinhead{Mesons}
$\pi^{\pm}$  & 0.1396 & 26.03                  & $2.21 \times 10^4$     & \checkmark\\
$K^{\pm}$    & 0.4937 & 12.39                  & 2970                   & \checkmark\\
$D^{\pm}$           & 1.869  & 0.00104                & 0.0660          & \checkmark\\
$D_s^{\pm}$         & 1.968  & $5.0 \times 10^{-4}$   & 0.030           & \checkmark\\
$B^{\pm}$           & 5.279  & 0.00164                & 0.0368          & \checkmark\\
$B_c^{\pm}$         & 6.286  & $5 \times 10^{-4}$     & 0.009           & \\
\cutinhead{Baryons}
$\Sigma^{\pm}$      & 1.189  & 0.0802                 & 8.00            & \\
$\Xi^-$             & 1.321  & 0.164                  & 14.7            & \\
$\Omega^-$          & 1.672  & 0.082                  & 5.8             & \\
$\Lambda_c^{\pm}$   & 2.286  & $2.00 \times 10^{-4}$  & 0.0104          & \checkmark
\enddata
\label{table:ChargedUnstable}
\tablecomments{Particle data as listed in \citet{Olive14}.  The typical distance a Planck energy particle travels before decaying is given as $\gamma_{\rm Planck} c\tau$.  A checkmark (\checkmark) appears under the ``Direct $\nu$/$\bar{\nu}$'' column if neutrinos are known to be among the immediate decay products in decay modes that occur more than 1\% of the time.}
\end{deluxetable}

\begin{deluxetable}{lcccc}
\tablecaption{Common unstable neutral particles}
\tablehead{\colhead{Name} & \colhead{Rest energy} & \colhead{Rest lifetime} & \colhead{$\gamma_{\rm Planck} c\tau$} & \colhead{Direct $\nu$/$\bar{\nu}$}\\ & \colhead{$(\GeV)$} & \colhead{$(\nsec)$} & \colhead{$(\pc)$}}
\startdata
$Z$        & 91.2    & $2.64 \times 10^{-16}$   & $3.43 \times 10^{-16}$         & \checkmark\\
$H$        & 125     & $\sim 2 \times 10^{-13}$ & $2 \times 10^{-13}$            & \\
\cutinhead{Mesons}
$\pi^0$    & 0.1350  & $8 \times 10^{-8}$       & $7 \times 10^{-5}$             & \\
$\eta$              & 0.5475  & $5 \times 10^{-10}$      & $1 \times 10^{-7}$    & \\
$K^0_S$    & 0.4976  & 0.0895                   & 21.3                           & \\
$K^0_L$    & 0.4976  & 51                       & $1.2 \times 10^4$              & \checkmark\\
$D_0$               & 1.865   & $4.10 \times 10^{-4}$    & 0.0214                & \checkmark\\
$B_0$               & 5.279   & 0.00153                  & 0.0344                & \checkmark\\
$B_s^0$             & 5.368   & 0.00147                  & 0.0325                & \checkmark\\
\cutinhead{Baryons}
$n$                 & 0.9396  & $8.86 \times 10^{11}$    & $1.1 \times 10^{14}$  & \checkmark\\
$\Lambda$           & 1.116   & 0.263                    & 27.9                  & \\
$\Sigma_0$          & 1.193   & $7.4 \times 10^{-11}$    & $7.4 \times 10^{-9}$  & \\
$\Lambda_b^0$       & 5.624   & 0.00123                  & 0.0259                & \checkmark
\enddata
\label{table:NeutralUnstable}
\tablecomments{Particle data as listed in \citet{Olive14}.  The typical distance a Planck energy particle travels before decaying is given as $\gamma_{\rm Planck} c\tau$.  A checkmark (\checkmark) appears under the ``Direct $\nu$/$\bar{\nu}$'' column if neutrinos are known to be among the immediate decay products in decay modes that occur more than 1\% of the time.}
\end{deluxetable}

We see most unstable particles decay before they can escape.  The minimum energy threshold for escape to win is
\begin{align}
\nonumber E_{\rm min}^{\rm esc} & = \frac{\Gamma^2 \delta t m c^2}{\tau} = 1\ \YeV\ \Gamma_3^2 m_{\GeV} \delta t_{\sec} \tau_{\nsec}^{-1}
\end{align}

\subsubsection{Synchrotron cooling}
The same electromagnetic fields that accelerate charged particles also induce synchrotron losses.  As long as the magnetic field isn't too high (in the Thomson regime), the cooling time $t_{\rm synch}^{\prime}$ decreases with energy as charged particles radiate faster.  The more rapid synchrotron cooling is in fact one of the main obstacles to building a Planck accelerator \citep{Thompson11}.  Indeed, synchrotron losses are one of the main limiting factors in current accelerators as well. 

The pitch-angle averaged synchrotron cooling time of a relativistic charged particle in the accelerator is
\begin{equation}
t_{\rm synch}^{\prime} = \frac{9}{32 \pi} \frac{(m c^2)^4}{Z^4 e^4 c E^{\prime} u_B^{\prime}}
\end{equation}
in the Thomson regime.  Technically, the emission mechanism could be ``diffusive synchrotron'' or ``jitter'' radiation if the magnetic field is very tangled on small scales, but the energy loss time is still the same \citep[e.g.,][]{Kelner13}.  If we assume that the magnetic energy density is $u_B^{\prime} = \epsilon_B u^{\prime}$, then the synchrotron cooling time is
\begin{align}
t_{\rm synch}^{\prime} = 22\ \nsec\ \frac{\Psi m_{\GeV}^4 \Gamma_3^7 \delta t_{\sec}^2}{Z^4 E_{28} \epsilon_B} \left(\frac{L}{L_E}\right)^{-1}.
\end{align}
For a beamed flow, the cooling time is
\begin{align}
t_{\rm synch}^{\prime} = 11\ \sec\ \frac{\psi_{\Gamma} m_{\GeV}^4 \Gamma_6^5 \delta t_{\sec}^2}{Z^4 E_{28} \epsilon_B} \left(\frac{L}{L_E}\right)^{-1}.
\end{align}

\begin{figure*}
\centerline{\includegraphics[width=8cm]{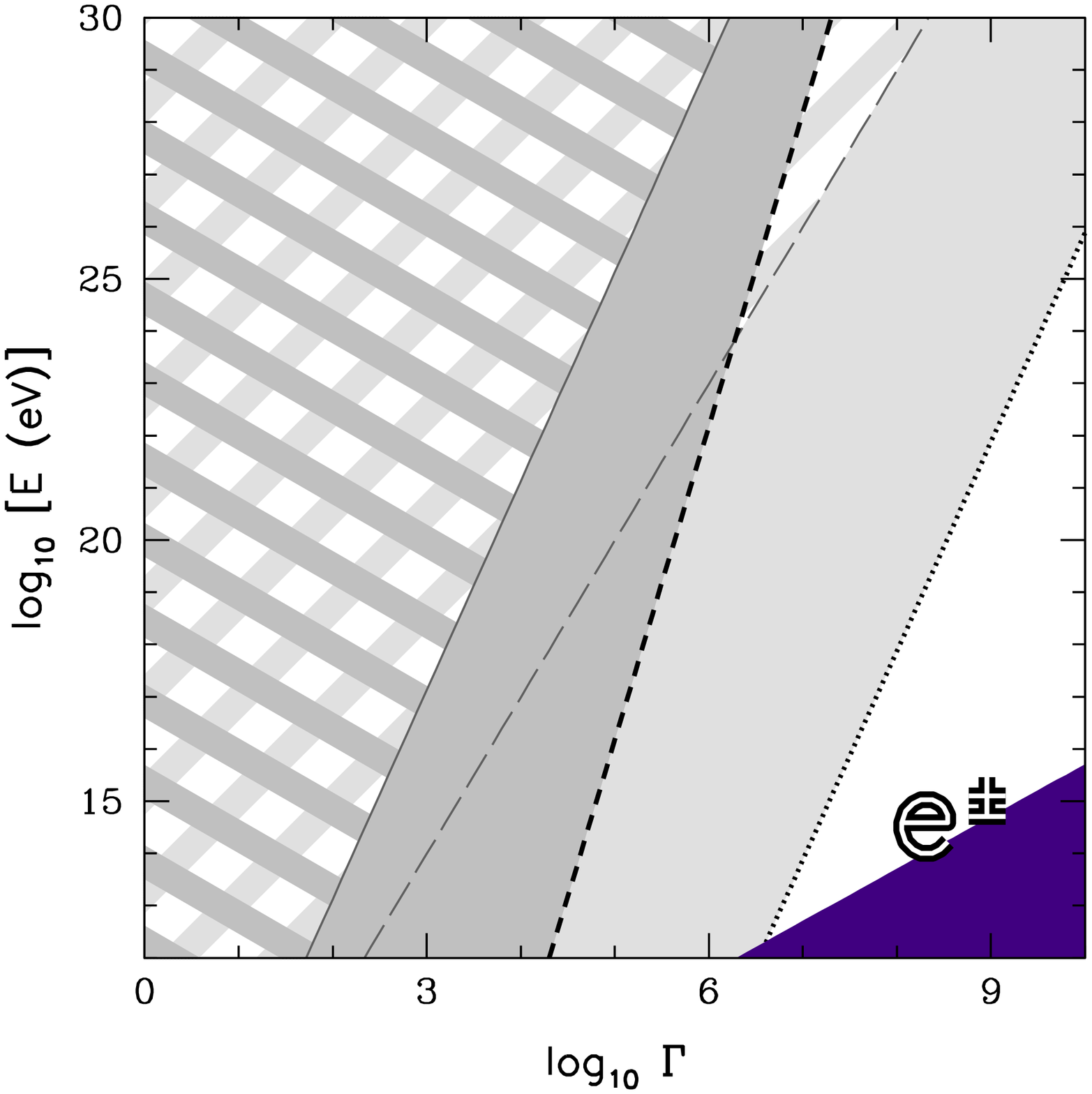}\includegraphics[width=8cm]{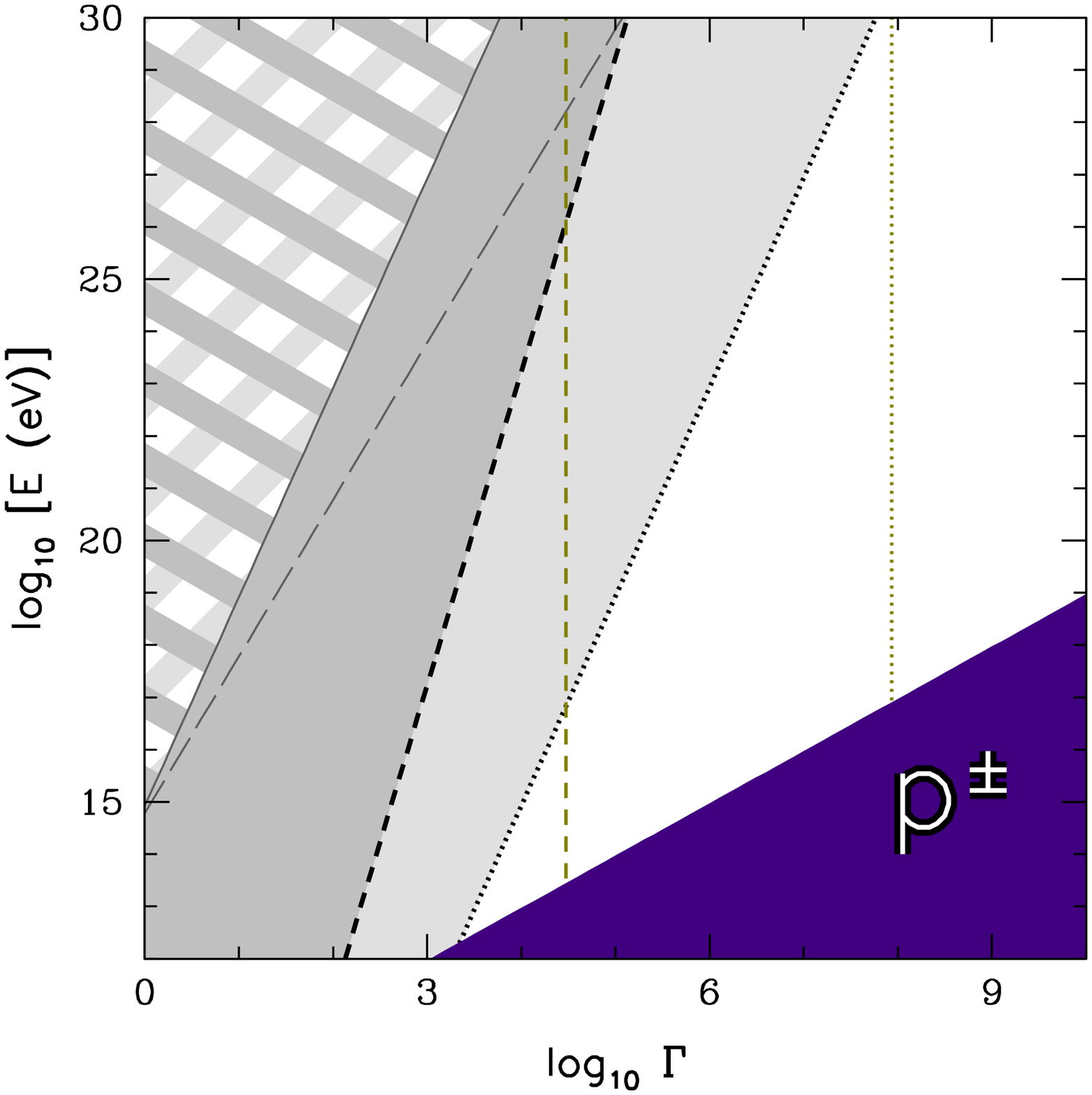}}
\figcaption{Synchrotron cooling regimes for the electron (left) and proton (right).  Particles in the dark grey shaded regions, left of the short-dashed lines, cool in isotropic outflows before they can escape.  For highly beamed outflows with $\psi_{\Gamma} = 1$, synchrotron cooling is efficient in the light grey shaded regions, left of the black dotted lines.  The particles are in the QED synchrotron cooling regimes in the cross-hatched regions, left of the grey solid (long-dashed) lines for isotropic (highly beamed) flows.  Protons left of the gold lines are stopped by thermal photons before escaping (short-dashed for isotropic flows; dotted for highly beamed flows).  The lower-right regions, shaded in dark blue, are below the boosted rest energies of the particles, and thus excluded.  I assume that $\delta t_{\sec} = \epsilon_B = \epsilon_{\rm rad} = \sigma_{\mbarn} = L / L_E = 1$.  \label{fig:SynchRegimesStable}}
\end{figure*}

Comparing to the dynamical time of the outflow, it is clear that the synchrotron cooling time is much shorter unless $\Gamma$ or $\delta t$ is very big.  Stable charged particles  cannot escape the accelerator region unless their engine-frame energy is less than 
\begin{align}
\nonumber E_{\rm synch}^{\rm esc} & = \frac{9}{8} \frac{\Psi c^2 \delta t \Gamma^6 (m c^2)^4}{Z^4 e^4 \epsilon_B L} = 220\ \PeV\ \frac{\Psi m_{\GeV}^4 \Gamma_3^6 \delta t_{\sec}}{Z^4 \epsilon_B} \left(\frac{L}{L_E}\right)^{-1} \\
                                  & = 110\ \ZeV\ \frac{\psi_{\Gamma} m_{\GeV}^4 \Gamma_6^4 \delta t_{\sec}}{Z^4 \epsilon_B} \left(\frac{L}{L_E}\right)^{-1}
\end{align}
Figure~\ref{fig:SynchRegimesStable} shows the energies and outflow $\Gamma$ where synchrotron cooling is important for electrons and protons in a fiducial maximal accelerator.  YeV electrons cool essentially instantly by synchrotron radiation, especially if the flow is beamed.  This demonstrates how hard it is to directly accelerate electrons to these energies, although they can be created as secondaries from other particles.  The situation is somewhat better for protons.  Planck energy protons can escape before cooling through synchrotron if $\Gamma \ga 3 \times 10^4\ (3 \times 10^7)$ for an isotropic (highly beamed) accelerator.

For unstable charged particles, synchrotron cooling is faster than particle decay if their energy is more than
\begin{align}
\nonumber E_{\rm synch}^{\rm decay} & = \Gamma^4 \sqrt{\frac{9}{8} \frac{\Psi c^2 \delta t^2 (m c^2)^5}{Z^4 e^4 \epsilon_B L \tau}}\\
\nonumber                           & = 460\ \ZeV\ \sqrt{\frac{\Psi m_{\GeV}^5}{\tau_{\nsec} \epsilon_B}} \frac{\Gamma_3^4 \delta t_{\sec}}{Z^2} \left(\frac{L}{L_E}\right)^{-1/2}\\
                                    & = 330\ \YeV\ \sqrt{\frac{\psi_{\Gamma} m_{\GeV}^5}{\tau_{\nsec} \epsilon_B}} \frac{\Gamma_6^3 \delta t_{\sec}}{Z^2} \left(\frac{L}{L_E}\right)^{-1/2}
\end{align}

\begin{figure*}
\centerline{\includegraphics[width=18cm]{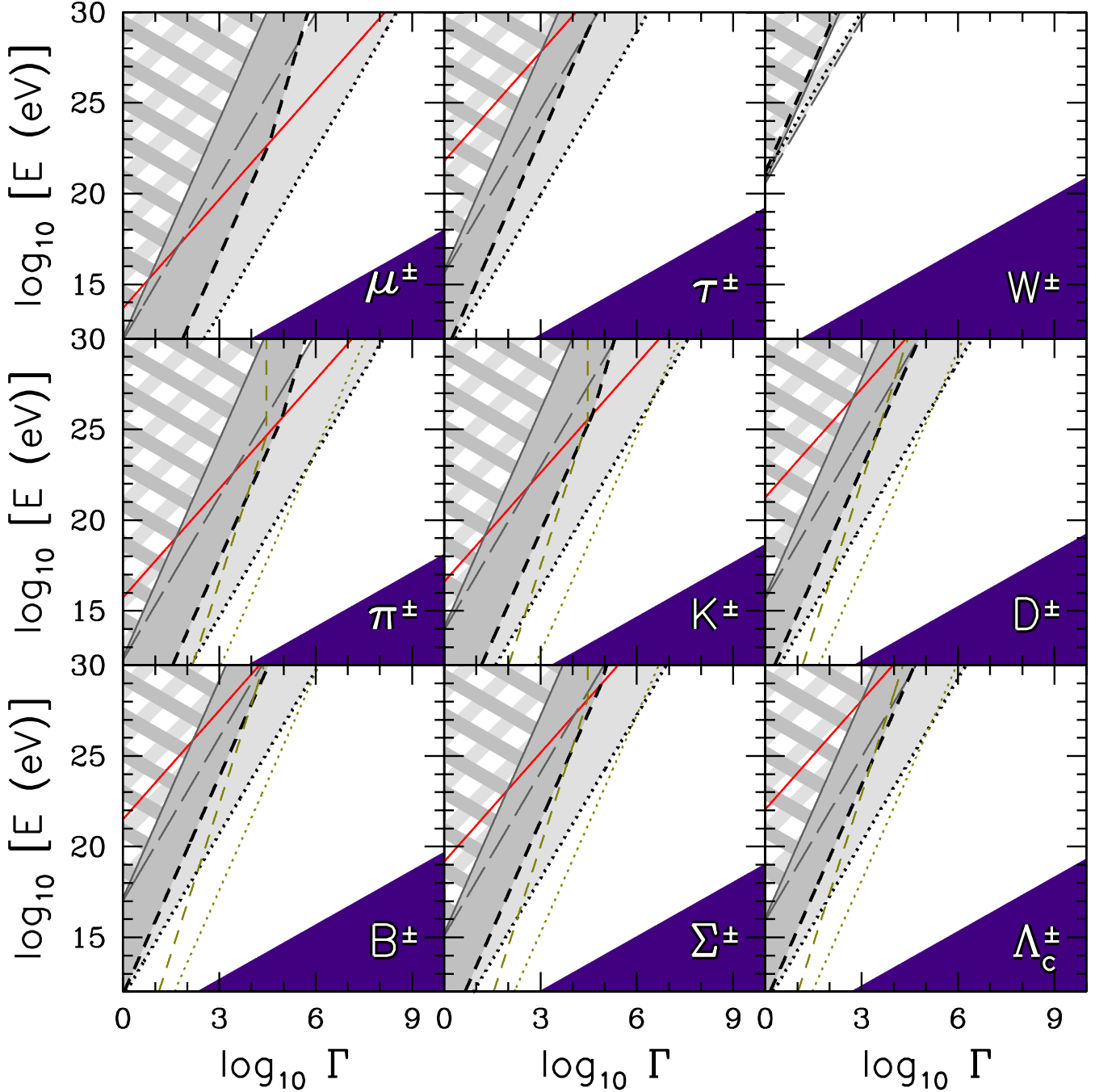}}
\figcaption{Synchrotron cooling regimes for various common unstable particles.  The shading and line styles are the same as in Figure~\ref{fig:SynchRegimesStable}.  In addition, the red solid lines indicate $E_{\rm min}^{\rm esc}$: below the line, the boosted particle lifetime is shorter than the escape time.  I assume that $\delta t_{\sec} = \epsilon_B = \epsilon_{\rm rad} = \sigma_{\mbarn} = L / L_E = 1$. \label{fig:SynchRegimesUnstable}}
\end{figure*}

These conditions are displayed in Figure~\ref{fig:SynchRegimesUnstable} for some commonly produced particles in accelerators.  The synchrotron losses are particularly bad for muons, which is important since they are a common decay product of hadrons and one of the main sources of neutrinos.  Mesons like pions are marginally better, since they have a shorter lifetime and larger masses (Table~\ref{table:ChargedUnstable}).  The heavier charmed ($D^{\pm}$, $D_s^{\pm}$) and beautiful mesons ($B^{\pm}$) and baryons ($\Lambda_c^{\pm}$), as well as the $\tau$ leptons, face less stringent losses.  These ``merely'' require outflows with $\Gamma \ga 10^4\ (10^5)$ to decay before cooling at Planck energy inside isotropic (extremely beamed) accelerators, emitting a hard component of prompt neutrinos \citep{Enberg09}.  Even the W bosons face synchrotron losses at Planck energy, unless the outflow is highly relativistic. 

\emph{The QED Limit} -- All of the above applies for particles in the Thomson regime.  This means that (1) the particle's motion is essentially classical, and the energy of emitted photons is much smaller than the particle's kinetic energy, and (2) the particle can be treated as a point charge \citep[e.g.,][]{Rybicki79}.  In the Weizs\"acker-Williams approach, synchrotron radiation is basically the scattering of a virtual photon by the Inverse Compton effect \citep{Lieu93}.  Much of the same physics that applies in scattering of photons also applies to synchrotron emission.

The classical condition is broken when $B^{\prime}$ is greater than $B_{\rm QED}^{\prime} = m^2 c^3 / (\hbar Z e \gamma^{\prime})$, or equivalently when the particles energy is bigger than
\begin{align}
\label{eqn:ESynchQED}
\nonumber E_{\rm synch}^{\rm QED} & = \frac{\sqrt{2} \pi \Psi c (m c^2)^3 \delta t \Gamma^4}{h Z e \sqrt{\epsilon_B L}}\\
\nonumber                         & = 1.0\ \XeV\ \frac{\sqrt{\Psi} m_{\GeV}^3 \Gamma_3^4 \delta t_{\sec}}{\sqrt{\epsilon_B} Z} \left(\frac{L}{L_E}\right)^{-1/2}\\
                                  & = 730\ \WeV\ \frac{\sqrt{\psi_{\Gamma}} m_{\GeV}^3 \Gamma_6^3 \delta t_{\sec}}{\sqrt{\epsilon_B} Z} \left(\frac{L}{L_E}\right)^{-1/2}
\end{align}
Then a calculation using QED is necessary.  In QED, the particle lies in discrete Landau energy states in the magnetic field, and emits photons when transitioning between those states.  For $E \gg E_{\rm synch}^{\rm QED}$ (the Klein-Nishina limit), the particle triggers an electromagnetic cascade, converting most of its energy into a photon.  Because of this, the particle's synchrotron cooling time actually increases as $E^{1/3}$ \citep[e.g.,][]{Harding06}.  But the problem becomes even more complicated when $B^{\prime} \ga \alpha_{\rm EM}^{-1} B_{\rm QED}^{\prime}$, and a full QED calculation is necessary \citep{Shen72,Nelson91,Lieu93}.

As far as we know, the leptons truly are point charges, but hadrons are composite particles with non-zero charge radii.  In the Weizs\"acker-Williams approach, the virtual photons' wavelengths become smaller than the hadrons themselves when the magnetic field is big enough, and the photons interact directly with constituent quarks.  As such, QCD processes become important in this regime, so there should be hadron emission as a result.  

The basic principle that particles emit synchrotron radiation in the form of any fields they interact with has been known for a long time, and the emission of synchrotron mesons has been calculated in simple approximations \citep{Ginzburg65}.  There have been semiclassical calculations of the production of mesons (pions, kaons, and eta) from protons emitting synchrotron \citep{Tokuhisa99,Herpay08-Synch,Kajino14} and curvature \citep{Berezinsky95,Herpay08-Curv,Fregolente12} radiation.  But no full QED/QCD calculation has been done, so the energy losses of hadrons are essentially unknown for $E \gg E_{\rm synch}^{\rm QED}$.

Equation~\ref{eqn:ESynchQED} is still the limit if the photon resolves the hadron when its wavelength in the hadron frame is shorter than the hadron's Compton wavelength.  The actual magnitudes of charge radii of hadrons are poorly known and discrepant, but are roughly a fraction of a fm, or $\sim 1\ \GeV$ in natural units \citep[e.g.,][]{Amendolia84,Amendolia86,Eschrich01,Hwang02,Pohl10}.  Since the hadrons I consider have masses of order $0.1$ -- $10\ \GeV$, I assume this is an adequate approximation.

The particle energies and outflow $\Gamma$ outside the Thomson regime are cross-hatched in Figure~\ref{fig:SynchRegimesStable} for electrons and protons and Figure~\ref{fig:SynchRegimesUnstable} for unstable particles.  It is conceivable that the synchrotron losses are slow enough in these regions that particles could escape or decay before cooling, but this awaits full QED and QCD calculations.  As we shall see, other losses probably intervene in these regions anyway.

\subsubsection{Baryonic cooling}
Aside from electromagnetic fields, an outflow can contain matter within it.  Thermal protons and neutrons within the accelerator region represent another obstacle for any particle trying to escape.  Escaping or decaying hadrons can be cooled by direct collisions with thermal nucleons, whether or not they are charged.  CR electrons and positrons lose their energy catastrophically to bremsstrahlung emission when they encounter baryons.  Likewise, opacity to $\gamma$-rays arises from thermal baryons due to pair production.  The typical grammage for all of these processes is several tens of $\gcm2$, which I will take to be $\Sigma_{\rm loss} = 50 \Sigma_{50}\ \gcm2$.

The density of baryons is found from the energy density (equation~\ref{eqn:uPrime}).  Let $\epsilon_{\rm bar} = \epsilon_{\rm bar} \epsilon$ be the energy density in baryons.  Then the baryonic density is
\begin{align}
\nonumber n_{\rm bar}^{\prime} = \frac{\epsilon_{\rm bar} L}{4 \pi \Psi \Gamma^6 (m_p c^2) c^3 \delta t^2} = 7.1 \times 10^{11}\ \cm^{-3} \frac{\epsilon_{\rm bar}}{\Psi \Gamma_3^6 \delta t_{\sec}^2} \left(\frac{L}{L_E}\right).
\end{align}

The grammage through the outflow's shell is $\Sigma^{\prime} = n_{\rm bar}^{\prime} m_N \delta r^{\prime}$.  By setting $\Sigma^{\prime}$ equal to $\Sigma_{\rm loss}$, I define a minimum $\Gamma$, below which particles are cooled by baryons:
\begin{align}
\nonumber \Gamma_{\rm bar}^{\rm esc} & = \left[\frac{\epsilon_{\rm bar} L}{4 \pi \Psi \Sigma_{\rm loss} c^4 \delta t}\right]^{1/5} = 940\ \left[\frac{\epsilon_{\rm bar}}{\Psi \delta t_{\sec} \Sigma_{50}} \left(\frac{L}{L_E}\right)\right]^{1/5} \\
                                     & = 1.1 \times 10^5\ \left[\frac{\epsilon_{\rm bar}}{\psi_{\Gamma} \delta t_{\sec} \Sigma_{50}} \left(\frac{L}{L_E}\right)\right]^{1/3}.
\end{align}
Baryonic cooling traps stable particles of all energies in slow outflows, but generally synchrotron losses are more important.

Unstable particles that would decay before escaping (see discussion in section~\ref{sec:EscapeVsDecay}) traverse only $\Sigma^{\prime} = c \gamma^{\prime} \tau n_{\rm bar}^{\prime} m_p = E c \tau n_{\rm bar}^{\prime} m_p / (\Gamma m)$.  The minimum $\Gamma$ in these cases are:
\begin{align}
\nonumber \Gamma_{\rm bar}^{\rm decay} & = \left[\frac{\epsilon_{\rm bar} L E \tau}{4 \pi \Psi \Sigma_{\rm loss} c^4 \delta t (m c^2)}\right]^{1/7} \\
\nonumber                              & = 3600\ \left[\frac{\epsilon_{\rm bar} E_{28} \tau_{\nsec}}{\Psi \delta t_{\sec} \Sigma_{50}} \left(\frac{L}{L_E}\right)\right]^{1/7} \\
                                       & = 1.1 \times 10^5\ \left[\frac{\epsilon_{\rm bar} E_{28} \tau_{\nsec}}{\psi_{\Gamma} \delta t_{\sec} \Sigma_{50}} \left(\frac{L}{L_E}\right)\right]^{1/5}.
\end{align}
This again confirms that synchrotron losses are probably more important.

Muons, $\tau$ leptons, and neutrinos are far more penetrative than other particles.  At Planck energies, their stopping length is $\Sigma \ga 10^6\ \gcm2$ \citep{IyerDutta01}.  Because of the small dependence of the critical $\Gamma$ on $\Sigma$, the estimates are not changed much for them.

\subsubsection{Cooling by thermal radiation}
\label{sec:ThermalRadiationCooling}
The intense activity in relativistic outflows can also generate a lot of heat.  The heat manifests mainly as photons.  The radiation energy density is some fraction $\epsilon_{\rm rad}$ of the total energy density.  If the heat radiation is thermal, it has a characteristic temperature
\begin{align}
\nonumber T^{\prime} & = \left[\frac{\epsilon_{\rm rad} L}{4 \pi \Psi \Gamma^6 a_{\rm SB} c^3 \delta t^2}\right]^{1/4} \\
\nonumber            & = 6.1 \times 10^5\ \Kelv\ \left[\frac{\epsilon_{\rm rad}}{\Psi \delta t_{\sec}^2 \Gamma_3^6} \left(\frac{L}{L_E}\right)\right]^{1/4}\\
                     & = 2.3 \times 10^4\ \Kelv\ \left[\frac{\epsilon_{\rm rad}}{\psi_{\Gamma} \delta t_{\sec}^2 \Gamma_6^4} \left(\frac{L}{L_E}\right)\right]^{1/4}.
\end{align}
Blackbody radiation has a number density $n_{\gamma}^{\prime} = u_{\gamma}^{\prime} / [\pi^4 k_B T^{\prime} / (30 \zeta(3))]  \approx u_{\gamma}^{\prime} / (2.7 k_B T^{\prime})$ \citep{Kolb90}, so the photon density in the fireball is
\begin{equation}
n_{\gamma}^{\prime} = \frac{30 \zeta(3)}{\pi^4 k_B} \left(\frac{a_{\rm SB} \epsilon_{\rm rad} L}{4 \pi \psi_{\Gamma} \Gamma^6 c^3 \delta t^2}\right)^{3/4}.
\end{equation}
This is generally $10^{14.5}$ -- $10^{18.5}\ \cm^{-3}$ for the parameters I have been using.  For leptons, the interaction with radiation results in Inverse Compton scattering, which behaves similarly to synchrotron cooling if the magnetic energy density is replaced by the radiation energy density.  Since the photon energy is a few eV and the particle energy is YeV scale, leptonic Inverse Compton is deep in the QED regime.  

Hadrons interact with the photons with some cross section $\sigma = 10^{-27} \sigma_{\mbarn}\ \cm^2$.  The cross section for hadrons interacting with photons is of order a mbarn \citep{Olive14}.  Generally, $\kappa \approx 0.1$ for these hadronic processes, so a hadron could survive multiple collisions with photons, but I ignore this factor since it extends the interaction time by only a factor of a few.  Stable particles escape if the optical depth $\Theta_{\rm esc} = n_{\gamma}^{\prime} \sigma \delta r^{\prime}$ is less than 1.  I find that the optical depth is
\begin{align}
\Theta_{\rm esc} & = 1.4 \times 10^5\ \left(\frac{\epsilon_{\rm rad} L}{L_E}\right)^{3/4} \Gamma_3^{-7/2} \delta t_s^{-1/2} \Psi^{-3/4} \sigma_{\mbarn}.
\end{align}
The fireball is clear enough to let particles escape only if $\Gamma$ is bigger than
\begin{align}
\nonumber \Gamma_{\gamma}^{\rm esc} & = \left[\left(\frac{30 \zeta(3) \sigma}{\pi^4 k_B}\right)^4 \left(\frac{a_{\rm SB} \epsilon_{\rm rad} L}{4 \pi \Psi}\right)^3 c^{-5} \delta t^{-2} \right]^{1/14}\\
\nonumber            & = 3.0 \times 10^4\ \left(\frac{\sigma_{\mbarn}^4 \epsilon_{\rm rad}}{\Psi^3 \delta t_s^2}\right)^{1/14} \left(\frac{L}{L_E}\right)^{3/14}\\
                     & = 8.6 \times 10^7\ \left(\frac{\sigma_{\mbarn}^4 \epsilon_{\rm rad}}{\psi_{\Gamma}^3 \delta t_s^2}\right)^{1/8} \left(\frac{L}{L_E}\right)^{3/8}.
\end{align}
The regions left of the gold lines in Figure~\ref{fig:SynchRegimesStable} are opaque.

The optical depth unstable hadrons experience over their lifetime is $\Theta_{\rm decay} = c n_{\gamma}^{\prime} \sigma E \tau / (\Gamma m c^2)$.  In order for photon scattering to be negligible, the Lorentz factor should be bigger than
\begin{align}
\nonumber \Gamma_{\gamma}^{\rm decay} & = \left[\left(\frac{30 \zeta(3) \sigma}{\pi^4 k_B} \frac{E \tau}{m c^2}\right)^4 \left(\frac{\epsilon_{\rm rad} L a_{\rm SB}}{4 \pi \Psi}\right)^3 c^{-5} \delta t^{-2} \right]^{1/22}\\
\nonumber            & = 4.6 \times 10^4\ \left(\frac{\sigma_{\mbarn}^4 \epsilon_{\rm rad}}{\Psi^3 \delta t_s^2}\right)^{1/22} \left(\frac{E_{28} \tau_{\nsec}}{m_{\GeV}}\right)^{2/11} \left(\frac{L}{L_E}\right)^{3/22}\\
                     & = 3.0 \times 10^6\ \left(\frac{\sigma_{\mbarn}^4 \epsilon_{\rm rad}}{\psi_{\Gamma}^3 \delta t_s^2}\right)^{1/16} \left(\frac{E_{28} \tau_{\nsec}}{m_{\GeV}}\right)^{1/4} \left(\frac{L}{L_E}\right)^{3/16}.
\end{align}

The thermal radiation probably is the biggest obstacle for the escape of protons and neutrons from the fireball, unless its luminosity is low or the particle energy is high.  That is simply because there is so many photons.  In the highly beamed case, it may be necessary for the fireball outflow to reach $\Gamma \ga 10^8$ for long-lived hadrons to escape.  The regimes of effective photon cooling of mesons and heavy baryons (gold lines in Figure~\ref{fig:SynchRegimesUnstable}) are fairly similar to the synchrotron cooling regimes (black lines), especially for Planck energy particles.  

\subsubsection{Summary}

The calculations in the previous subsections all basically highlight the compactness problem.  When an astrophysical explosion contains a lot of energy, the fireball is so dense that it is opaque unless it is highly relativistic \citep{Cavallo78}.  If Planck accelerators need luminosities comparable to the Einstein luminosity, the compactness problem is especially severe.  Although outflows with $\Gamma$ of several hundred are actually achieved in GRBs \citep{Lithwick01,Abdo09-GRB}, even that may not be enough to release Planck energy particles, with $\Gamma \ga 10^5$ required for the fireball to be sufficiently transparent for radiation to escape.

Although the internal shocks model I use is applicable to astrophysical transients, it assumes that the accelerator is very chaotic and inefficient.  Basically, a fireball model describes a bomb being set off, with the ETs presumably just watching the particles that it produces.  An accelerator built by ET may be much ``cleaner'', with much less energy density.  After all, the ETs themselves need to observe the collisions.  Arguably both approaches have precedent in our own history.  Traditional particle accelerators like the LHC are very clean, so that the events can be analyzed easily.  But in the 1950s and 1960s, nuclear weapons were detonated in space to test the effects of energetic particles in the Earth's magnetosphere (these tests thankfully have ceased).

\subsection{Conditions between the accelerator and Earth}
\label{sec:Propagation}
Even if it does escape the accelerator, YeV radiation still does not necessarily reach us.  In order to do that, it must (at least) escape from the accelerator's host galaxy (if any), traverse intergalactic space, and then travel within the Milky Way to Earth.  Particles can be blocked from reaching Earth if they decay before making it here, if they cool radiatively, or if they interact with photons.

The magnetic fields and the extragalactic radiation fields in the IGM and the Milky Way are the biggest obstacles for YeV particles.  I take the Milky Way to have a magnetic field of $B = 6\ \muGauss$, which extends over 10 kpc \citep{Strong00,Beck01}.  Presumably our Galaxy is typical of galaxies hosting ETs.  But note that at $z \approx 1$--2, there was a population of luminous infrared main-sequence galaxies that contained most star formation.  These probably had magnetic fields larger than the Milky Way's, perhaps of order $\sim 100\ \muGauss$ to $1\ \mGauss$ \citep{Lacki10-zHiFRC}, judging from the fact that they lie on the far-infrared--radio correlation observed for star-forming galaxies \citep[e.g.,][]{Mao11}.  On the other hand, ETs may live in quiescent, red galaxies, which probably have somewhat weaker magnetic fields than the Milky Way \citep{Moss96}.  

Unlike CRs at ZeV and lower energies, particles at YeV energies are not deflected significantly by magnetic fields.  Their gyroradii are
\begin{equation}
R_L = 180\ Z^{-1}\ \Mpc\ \left(\frac{E}{1\ \YeV}\right) \left(\frac{B}{6\ \muGauss}\right)^{-1}.
\end{equation}
Even if the Galaxy's magnetic field is perfectly regular, it deflects a YeV particle by at most $R_{\rm MW} / R_L \approx 11\farcs Z (E/\YeV)^{-1}$.  The intergalactic magnetic field is very poorly known, but I scale to $B_{\rm IGM} = 1\ \nGauss$ with coherence lengths of $\ell_{\rm IGM} = 1\ \Mpc$.  The average deflection of a YeV CR is $\sqrt{D / \ell_{\rm IGM}} (\ell_{\rm IGM} / R_L) = \sqrt{D \ell_{\rm IGM}} Z e B_{\rm IGM} / E \approx 6\farcs Z (D / \Gpc)^{1/2} (\ell / \Mpc)^{1/2} (B_{\rm IGM} / \nGauss) (E / \YeV)^{-1}$.  Thus, even charged particles should point directly back at their sources to high precision.  The limiting factor on the angular resolution is the ability of experiments to reconstruct the trajectory of the CR.  

By far, most extragalactic photons are part of the Cosmic Microwave Background (CMB), with a number density $n_{\rm CMB} = 411\ \cm^{-3}$ \citep{Olive14}.  The extragalactic background light (EBL) from galaxies at ultraviolet to infrared wavelengths has a much smaller energy density than the CMB \citep[e.g.,][]{Franceschini08}, and each photon contains much more energy.  Nor is the radio background a problem, whether in the Galaxy or outside of it.  The Galactic radio brightness temperature peaks at $\sim 10^7\ \Kelv$ around 2 MHz \citep{Brown73-MHz,Fleishman95,Manning01}, implying a photon number density of $\sim 1\ \cm^{-3}$.  Strictly speaking, the extragalactic radio background is very poorly constrained at frequencies below 1 MHz, with brightness temperatures of $\gg 10^{12}\ \Kelv$ possible \citep{Lacki10-SubMHz}.  The expected radio background at these frequencies, however, is thought to be quite small \citep{Protheroe96-Radio}.

\subsubsection{Could we directly detect beam nucleons?}
\label{sec:NucelonRadiation}
Planck accelerators are likely to produce a lot of pollution when beam particles collide with each other, according to the arguments in Section~\ref{sec:BrightAccelerators}.  But maybe not all beam particles are consumed by these collisions, with some escaping the accelerator.  Perhaps the acceleration process is highly random, with only some of the CRs being ``harvested'' for collisions.  Another possibility is that accelerated particles are released between experimental runs.  In our colliders, beam particles are directed into a beam dump when the experiment is over.  The beam carries a lot of energy, which poses engineering challenges \citep{Schmidt06}.  With a Planck accelerator, the amount of energy in the beams is literally cosmic, so a beam dump may not be practical.  Instead, the beam particles might simply be released into interstellar space.  As noted in Section~\ref{sec:ThermalRadiationCooling}, the collider must be clear of thermal photons in order for beam nucleons to escape.

Nucleons experience the Greisen-Zatsepin-Kuz'min (GZK; \citealt{Greisen66,Zatsepin66}) effect as they propagate over intergalactic distances, in which photohadronic reactions occur with CMB photons.  The effect occurs at energies above 40 EeV, and has a typical energy loss scale of $\sim 20\ \Mpc$.  This GZK ``horizon'' for UHECRs is well known \citep{Bhattacharjee00}.  The reactions produce pions, which decay into neutrinos, $e^{\pm}$, and $\gamma$-rays.  The resulting population of GZK neutrinos are themselves sought in many experiments (section~\ref{sec:NeutrinoPropagation}).

In addition, protons at YeV energies experience synchrotron cooling in any intergalactic magnetic fields.  Nanogauss magnetic fields cool protons over a timescale $t_{\rm synch} = 140\ \Tyr\ (E/\YeV)^{-1} (B / \nGauss)^{-2}$, which conceivably matters for Planck energy protons.  If the intergalactic magnetic fields are picogauss or lower, as often thought, then synchrotron cooling poses no problem.

Protons traversing the Galactic magnetic field have a cooling time of $t_{\rm synch} = 4\ \Myr\ (E/\YeV)^{-1} (B / B_{\rm MW})^{-2}$.\footnote{They are in the Thomson regime at YeV--WeV energies.  QED effects only matter when $\gamma \la B_{\rm QED} / B_{\rm MW} \approx 3 \times 10^{25}$.}  Thus, synchrotron cooling prevents protons with energies above $\sim 100\ \YeV$ from reaching Earth without severe energy losses.  Instead, most of the CR proton energy would be converted into photons with typical energies $\gamma^2 h e B / (m c) \approx 400\ \eV\ (E / \YeV)^2 (B / B_{\rm MW})$ \citep{Rybicki79}.  Neutrons are not synchrotron cooled, and they are long-lived enough at YeV energies to reach us even from across the Universe.  The lifetime of a neutron is $\gamma \tau = 30\ \Gyr\ (E / \YeV)$; they are stable for our purposes.  They are stopped by the GZK effect, though.

While it takes a column of $\sim 80\ \gcm2$ to stop a GeV proton in hydrogen, the inelastic nucleon-nucleon cross section increases with energy.  When a YeV nucleon hits a proton at rest, $\sqrt{s} \approx 30\ \PeV\ (E / \YeV)^{1/2}$, and the approximate inelastic hadron-hadron collision cross section is $\sim 300\ \mbarn$ \citep{Olive14}.  In other words, it takes $\sim 5\ \gcm2$ to stop a YeV proton in hydrogen, a column attained only in the most extreme starburst regions and in Compton-thick AGNs.  The Milky Way's gas poses no obstacle for these CRs.

\subsubsection{Unstable hadrons and leptons}
\label{sec:MesonRadiation}
Hadronic colliders make a vast number of mesons, which are another possible type of radiation from Planck accelerators.  Mesons in flight frequently decay into muons, with a lifetime of a few microseconds.  The lifetimes of these unstable particles are time dilated so that $c\gamma \tau = 0.2\ \pc\ (E / \YeV) (\tau / 10\ \nsec)$ on average (Tables~\ref{table:ChargedUnstable} and \ref{table:NeutralUnstable}).  The longest-lived neutral mesons, particularly the $K_0^L$, might propagate a distance of a few kiloparsecs from their sources.

Charged mesons and muons experience synchrotron losses before they can decay, however, limiting their ranges to
\begin{align}
\nonumber c t_{\rm max} & = \left[\frac{9}{4} \frac{(m c^2)^3 \tau c}{Z^4 e^4 B^2}\right]^{1/2} = 0.1\ \kpc\ \sqrt{m_{\GeV}^3 \tau_{\nsec}} \left(\frac{B}{6\ \muGauss}\right)^{-1}.
\end{align}
The maximum range is attained when the particle has an energy
\begin{equation}
E_{\rm furthest} = 1\ \XeV\ \sqrt{\frac{m_{\GeV}^5}{\tau_{\nsec}}} \left(\frac{B}{6\ \muGauss}\right)^{-1}.
\end{equation}
Muons with an energy of 1 YeV have a range of 200 pc in the Galaxy.  Charged pions only make it 30 pc (at 20 YeV), while kaons may be able to go 150 pc at 600 YeV.  The heavier hadrons listed in Table~\ref{table:ChargedUnstable} all have ranges of less than 100 pc, reached only when their energy is many XeV.  

Given the apparent lack of cosmic engineering in the Solar neighborhood, it is unlikely that these particles can be directly detected at Earth. 

\subsubsection{Electromagnetic cascades from $e^{\pm}$ and $\gamma$-rays}

Although there are many ways of generating $\gamma$-rays and $e^{\pm}$ in colliders, including annihilation, pair production, and meson decay, they also strongly couple to the radiation and magnetic backgrounds of the IGM and the Galaxy.  Photons convert into $e^{\pm}$ when they annihilate with background photons, while $e^{\pm}$ convert into photons when they cool by Inverse Compton scattering, synchrotron, or triplet pair production (TPP; \citealt{Mastichiadis91,Anguelov99}).  During each conversion, the particle keeps most but not all of its energy.  Thus, over many conversions, the particle's energy cascades into lower energy forms.  Conventionally, it is thought that any energy in ultra high energy (UHE) photons or $e^{\pm}$ cascades down into MeV--GeV $\gamma$-rays over Gpc distances.  The unresolved $\gamma$-ray background then constrains the amount of power being generated in UHE $\gamma$-rays and $e^{\pm}$ in the Universe \citep{Coppi97,Murase12}.

There are some subtleties, though, when the primary $\gamma$-ray or $e^{\pm}$ energy is YeV or above.  At these energies, $\gamma$-rays annihilate with CMB photons through the double pair production process, with an interaction length of 120 Mpc \citep{Bhattacharjee00,Protheroe96-Cascades}.  YeV--XeV $e^{\pm}$ interact with the CMB through TPP.  But while the TPP cross section is large, its inelasticity is extremely small, $\la 10^{-7}$, so the energy attenuation length of $e^{\pm}$ is $>10\ \Gpc$ for these particles \citep{Bhattacharjee00,Protheroe96-Cascades}.  

Cascades can still proceed at these energies through synchrotron cooling, if there are magnetic fields anywhere along the line-of-sight to the source.  Picogauss magnetic fields cause electrons to lose energy in just $t_{\rm synch} = 13\ \yr\ (E / \YeV)^{-1} (B / \nGauss)^{-2}$.  While the volume-filling intergalactic magnetic field might be as small as $10^{-19}\ \Gauss$ \citep{Neronov10,Dermer11}, the necessary picogauss to nanogauss magnetic fields may be present in large-scale structures like galaxy filaments \citep[e.g.,][]{Dolag05}.  An $e^{\pm}$ traveling over cosmic distances is likely to hit one of these filaments and then cool.  There are probably also femtogauss magnetic fields generated by thermal fluctuations throughout the IGM \citep{Schlickeiser12,Yoon14}.  The synchrotron $\gamma$-rays emitted by the $e^{\pm}$ have characteristic energies of $0.5\ \keV\ (E / \YeV)^2 (B / \nGauss)$.

Any YeV photon that gets near the Milky Way is destroyed by pair production on the magnetic field, $\gamma + B \to e^+ + e^-$ \citep{Stecker03}.  The threshold for this process is 
\begin{equation}
E_{\rm thresh} = \frac{{\cal B}_{\rm QED}^e}{B_{\rm MW}} m_e c^2 = 4\ \YeV\ \left(\frac{B_{\rm MW}}{6\ \muGauss}\right)^{-1}.  
\end{equation}
This initiates an electromagnetic cascade.  At 1 YeV, the cooling time for $e^{\pm}$ is just 10 seconds in the Thomson regime; higher energies enter the QED regime, but the cooling time increases slowly with energy.  Thus, the energy in YeV--XeV photons or $e^{\pm}$ downgrades into ZeV photons within the Galaxy.  These can be directly detected by CR arrays like Auger.  So far, no showers from photons have been detected, setting a very stringent limit on their flux, much more stringent than the GeV $\gamma$-ray background implies \citep{Rubtsov06,Abraham09}.

Clearly, what happens to YeV--XeV photons and $e^{\pm}$ needs to be studied in more detail, but they have two possible fates.  If they come from distant extragalactic sources and there are femtogauss--picogauss magnetic fields along the line of sight, they cascade into MeV--GeV synchrotron $\gamma$-rays \citep{Coppi97,Murase12}.  Otherwise, they cascade into ZeV photons when reaching the Galaxy.

\subsubsection{Neutrinos}
\label{sec:NeutrinoPropagation}
The most promising way to detect Planck accelerators is almost certainly neutrinos.  These may be created in the collider itself through hadronic processes, or through the GZK process as free protons interact with the CMB.  It is even possible that the neutrinos are the beam particles themselves.  The extreme environments of accelerators might cool particles that produce neutrinos, but a small fraction of the power accelerated can still leak out as ``prompt'' neutrinos emitted by very quickly decaying hadrons with charm and beauty quarks \citep{Enberg09}.  In fact, neutrinos may be the only high energy particle that escapes the accelerator.  

YeV--XeV neutrinos propagate freely through the Universe.  Although they can interact with the CMB or the relic neutrino background \citep{Roulet93}, the Universe has an optical depth of only $\sim 0.001$ \citep{Seckel98}.  They can also cascade in magnetic fields \citep{Kuznetsov97}, but the threshold for this process is
\begin{equation}
E_{\rm thresh} = 150\ \VeV\ \left(\frac{B}{6\ \muGauss}\right).
\end{equation}
for electron neutrinos and even higher for muon and tau neutrinos \citep{Bhattacharya09}.  Even in the Earth's magnetic field ($B \approx 1\ \Gauss$), electron neutrinos are not affected unless they have energies of $\ga 100\ \XeV$.  In this case, the attenuation length is $2 \times 10^{15}\ \cm\ (B / \Gauss)^{-2} (E / 100\ \XeV)^{-1}$ \citep{Erdas03}, much bigger than the Earth itself except at the most extreme energies.

\subsection{Should we expect to see anything?}
\label{sec:HiddenAccelerators}
A Planck accelerator is literally a cosmic investment by anyone who builds it.  The amount of effort required to build and operate one, and the rarity of Planck events, could push ETs to be as efficient as possible in extracting information from it.  They may want to carefully study \emph{every} outgoing particle made in the collisions.  Would they let any escape so that we could detect them?  There could also be environmental reasons to stop high energy radiation, to prevent it from harming life elsewhere in the host galaxy.

Our own accelerators frequently measure the energy of outgoing particles with calorimeters, which stop the particles entirely.  One could imagine an ET somehow building calorimeters extreme enough to stop even neutrinos, or at least stopping mesons and leptons before they can decay into neutrinos.  However, calorimeters are not necessary to characterize the collisions.  Outgoing charged particles emit radiation and ionize matter, which can provide a signature of their presence (chapter 33 of \citealt{Olive14} is an extensive review of particle detectors).

We have some reasons to hope that ETs would at least allow neutrinos to escape the accelerator:
\begin{itemize}
\item It takes an enormous column to stop a YeV--XeV neutrino, about $\sim 10^6\ \gcm2$, and Planck accelerators are necessarily big.  A neutrino-stopping spherical shell with radius given by equation~\ref{eqn:AcceleratorSize} would have a mass of $0.005\ \Msun$, and that is with $B_{\rm QED}$ electromagnetic fields in the accelerator.  A Planck accelerator can easily be the size of a Solar System, in which case a spherical neutrino-stopping shell requires $\ga 10^4\ \Msun$ of material.  Of course, these requirements are a lot less burdensome if outgoing neutrinos are beamed.

\item The energy dumped in calorimeters does not simply disappear, but is dissipated.  This presents a lot of practical challenges.  There would be a large thermal background in the detectors from the heat generated, which could make detecting individual particles more difficult.  Stopped particles ionize atoms in the detectors as well, causing damage.  Finally, the calorimeters would become radioactive under the onslaught of radiation.  Aside from the problem of disposing of cosmic amounts of radioactive waste, the high energy emission from radioactive nuclei can mask desired signals.

\item A Planck accelerator also potentially serves as a YeV--XeV neutrino factory.  The propagation of YeV--XeV neutrinos over interstellar distances sets very tight constraints on new physics like Lorentz Invariance Violation.  These effects are expected to appear around the quantum gravity scale.  Our own precision tests can only constrain low-order effects with weak energy dependence \citep[e.g.,][]{Abdo09-LIV}, but this is not a problem for Planck energy neutrinos.  If the neutrinos are released for experiments elsewhere in their host galaxies to detect and characterize, we too could detect them at extragalactic distances.  Altruistic ETs might even build their Planck accelerator to be used as a cosmic neutrino factory observed throughout the Universe by other ETs.

\item Most speculatively, the ETs might release YeV--XeV neutrinos to advertise their having a Planck accelerator \citep[c.f.,][]{Swain06}.  Of course, this assumes that ETs want to communicate.  But if ETs generally find cosmic engineering distasteful, a few may build Planck accelerators and then draw attention to it to dissuade others from building their own.
\end{itemize}

But ultimately if all radiation is trapped by the accelerator, then it will appear to us only as waste heat.  Cosmic engineering can generally be detected by its waste heat in the infrared \citep{Dyson60,Wright14-Results}.  An artificial Planck accelerator may be much hotter than a structure like a Dyson sphere, though, since it is not primarily intended to be lived on.  On the one hand, it could be a NIR-excess source, looking like an obscured AGN.  In fact, it could actually \emph{be} a rare subtype of obscured AGN, especially if it uses black holes \citep[c.f.,][]{Kardashev95,Thompson11}.  But if it takes the form of a relativistic fireball (section~\ref{sec:AcceleratorConditions}), most of its thermal radiation could emerge as X-rays or soft $\gamma$-rays.  These accelerators would be transient, appearing more like a GRB than anything else, and would be missed by our current surveys for cosmic engineering.

\section{Y\lowercase{e}V neutrino limits}
Methods for detecting high energy neutrinos are basically the same as for detecting other types of UHECRs.  An UHE neutrino has some probability of colliding with the particles (typically nuclei) in a target volume.  If an interaction does occur, a shower of particles is produced.  Most of these particles, including hadrons, photons, electrons, muons, and $\tau$ leptons, emit detectable signals as they propagate from the collision.  An experiment can watch for these signals in the target volume.

High energy neutrinos are easier to detect than low energy neutrinos because the cross section for weak interactions is greater at larger energies.  At most energies, neutrinos are more likely to interact with nuclei than electrons, with a nucleon-neutrino cross section of 
\begin{equation}
\sigma_{\nu N} \approx 2 \times 10^{-30}\ \cm^{-2} \left(\frac{E_{\nu}}{\YeV}\right)^{0.363}
\end{equation}
from \citet{Gandhi98}.  The typical column to stop an UHE $\nu$ is thus
\begin{equation}
\label{eqn:UHENuColumn}
\Sigma_{\nu N} = \frac{m_p}{\sigma_{\nu N}} \approx 8 \times 10^5\ \gcm2\ \left(\frac{E_{\nu}}{\YeV}\right)^{-0.363},
\end{equation}
which is equivalent to about 8 km of ice or water, or about 1 or 2 km of rock.

In addition to a greater interaction probability, higher energy neutrinos that do interact produce a louder cascade signal.  The neutrino emits a weak boson carrying about 20\% of its energy, which decays into a cloud of hadrons \citep{Scholten06}.  The hadrons produce Cherenkov radiation in the medium that can be detected by an experiment.  The amount of Cherenkov radiation emitted is directly proportional to the initial energy of the UHE neutrino.  The Cherenkov radiation of the highest energy neutrinos can be observed from a great distance, so detectors for these neutrinos do not have to be sensitive at all.

Unfortunately, since more energy is packed into each neutrino, the number flux of YeV neutrinos is likely very low even if the energy flux is reasonably high.  The target area needed to catch even one neutrino at these energies must therefore be huge.  This is reflected in the bigger target areas for UHE neutrino detection experiments -- from the cubic kilometer IceCube at PeV \citep{Gaisser14}, to $\sim 200\ \km^2$ for the Askaryan Radio Array (ARA) at EeV energies \citep{Allison12}, to $\sim 10^5\ \km^2$ for JEM-EUSO at ZeV energies \citep{Takahashi09}, to the entire surface of the Moon's near side in the NuMoon lunar Cherenkov experiments \citep{Buitink10}.  The largest target that we can expect to observe in the near future is the entire Earth's surface area.  I define a unit of \emph{Earths} for a YeV neutrino experiment:
\begin{equation}
1\ \oplus = 4 \pi R_{\oplus}^2 = 5.11 \times 10^8\ \km^2.
\end{equation}

\subsection{Current and near-future experiments}
\label{sec:CurrentExperiments}
Any given YeV neutrino is relatively easy to detect when it interacts with the target volume; the challenge is the rarity of these events.  In Figure~\ref{fig:CurrentLimits}, I show the \emph{number flux} limits on UHE neutrinos from various experiments.  The literature containing these results generally do not extend the resulting constraints to XeV energies.  But as I noted, a neutrino becomes easier to detect at higher energy, emitting a bigger cascade signal.  I therefore extend the number flux limits to YeV--XeV energies in two ways: (1) by assuming the sensitivity is constant at the highest energies (dashed), and (2) assuming the sensitivity has the $E^{-0.363}$ scaling of the neutrino-nucleon cross section (dotted; equation~\ref{eqn:UHENuColumn}).

The relevant quantity is the effective etendue ${\cal A}$ of the experiment.  This is the product of the effective field of view $\Omega_{\rm eff}$ and the effective area $a_{\rm eff}$ of the detector (which is the actual area of the target scaled by the detector efficiency).  The number of expected events is then
\begin{equation}
\label{eqn:NEventsIsotropic}
\mean{N_{\rm events}} = E \frac{d\phi}{dE} {\cal A} t_{\rm exp},
\end{equation}
where $t_{\rm exp}$ is the livetime of the experiment.  From the lack of events ($\mean{N_{\rm events}} \la 1$) and the flux limits given by each experiment, I calculate their ${\cal A}$.  These are listed in Table~\ref{table:YeVExperiments} for the maximum energies $E_{\rm best}$ the bounds are given.  Since $E_{\rm best}$ varies over many orders of magnitude, I estimate ${\cal A}$ for every experiment at 100 YeV by scaling with the neutrino-nucleon cross section energy dependence \citep{Gandhi98}.

\begin{deluxetable*}{lccccccccc}
\tablecaption{UHE neutrino experiments}
\tablehead{\colhead{Experiment\tablenotemark{a}} & \multicolumn{4}{c}{Epochs} & \multicolumn{4}{c}{Best reported etendue} & \colhead{Extrapolated\tablenotemark{b}}\\ & \colhead{Start Date} & \colhead{End Date} & \colhead{$\Delta t$} & \colhead{$t_{\rm exp}$} & \colhead{$E_{\rm best}$} & \colhead{Best $E\,d\phi/dE$ limit} & \multicolumn{2}{c}{${\cal A}_{\rm best}$} & \colhead{${\cal A}\,(100\ \YeV)$} \\ & & & \colhead{$(\yr)$} & \colhead{$(\yr)$} & \colhead{$(\YeV)$} & \colhead{$(\km^2\,\sr^{-1}\,\yr^{-1})$} & \colhead{$(\km^2\,\sr)$} & \colhead{$(\oplus\,\sr)$} & \colhead{$(\km^2\,\sr)$}}
\startdata
Auger\tablenotemark{c} & 2004        & 2010 May 31 & 6     & 3.5    & $10^{-4}$ & 24                   & 0.012             & $2.3 \times 10^{-11}$ & 2               \\
RICE (2010)      & 1999        & 2010 Dec 31 & 12    & 5.3    & 0.001     & 0.51                 & 0.38              & $7.4 \times 10^{-10}$ & 30              \\
RICE (2005)      & 1999        & 2005 Aug 15 & 7     & 2.4    & 0.01      & 0.79                 & 0.54              & $1.1 \times 10^{-9}$  & 20              \\
LUNASKA (No SSR) & 2008 Feb 26 & 2006 May 19 & 0.23  & 0.0030 & 0.1       & 0.63                 & 530               & $1.0 \times 10^{-6}$  & 7000            \\
\phantom{LUNASKA }(SSR) &      &             &       &        &           & 0.032                & $1.1 \times 10^4$ & $2.1 \times 10^{-5}$  & $1 \times 10^5$ \\
ANITA-II         & 2008 Dec 21 & 2009 Jan 22 & 0.088 & 0.078  & 0.1       & 0.0082               & 1600              & $3.1 \times 10^{-6}$  & $2 \times 10^4$ \\
RESUN            & 2009 Aug    & 2009 Dec    & 0.3   & 0.023  & 1         & 0.02                 & 2200              & $4.3 \times 10^{-6}$  & $1 \times 10^4$ \\
SAUND II         & 2006 Jul    & 2007 Sep    & 1.2   & 0.36   & 3         & 9.5                  & 0.30              & $5.8 \times 10^{-10}$ & 1               \\
WSRT / NuMoon    & 2007 Jun 9  & 2008 Nov 11 & 1.4   & 0.0053 & 8.5       & $6.9 \times 10^{-5}$ & $2.7 \times 10^6$ & 0.0053                & $7 \times 10^6$ \\
FORTE            & 1997 Sep    & 1999 Dec    & 2.3   & 0.0082 & 100       & $9.5 \times 10^{-5}$ & $1.3 \times 10^6$ & 0.0025                & $1 \times 10^6$ \\
\cutinhead{Future experiments}
ARA              & \nodata     & \nodata     & 3.0   & $\sim 3$ & $3 \times 10^{-4}$ & 0.40                 & 0.83  & $1.6 \times 10^{-9}$ & 84\\
JEM-EUSO         & \nodata     & \nodata     & 5.0   & $\sim 5$ & 0.1                & $8.2 \times 10^{-4}$ & 240   & $4.8 \times 10^{-7}$ & 3000\\
LOFAR            & \nodata     & \nodata     & \nodata & 0.019  & 7.4                & $5.5 \times 10^{-6}$ & $9.5 \times 10^6$ & 0.019    & $2 \times 10^7$\\
SKA1-Low         & \nodata     & \nodata     & \nodata & 0.11   & 4.3                & $5.6 \times 10^{-6}$ & $1.6 \times 10^6$ & 0.0031   & $5 \times 10^6$
\enddata
\label{table:YeVExperiments}
\tablenotetext{a}{I used the following references. {\bf Auger}: \citet{Abreu12}. {\bf RICE (2010)}: \citet{Kravchenko12}.  {\bf RICE (2005)}: \citet{Kravchenko06}.  {\bf LUNASKA}: \citet{James10}.  {\bf ANITA}: \citet{Gorham10,Gorham12}.  {\bf RESUN}: \citet{Jaeger10}.  {\bf SAUND II}: \citet{Kurahashi10}.  {\bf WSRT / NuMoon}: \citet{Buitink10}.  {\bf FORTE}: \citet{Lehtinen04}.  {\bf ARA}: \citet{Allison12}.  {\bf JEM-EUSO}: \citet{Ebisuzaki11,Medina-Tanco11}.  {\bf LOFAR}: \citet{Buitink13}.  {\bf SKA1-Low}: \citet{Bray14}.}
\tablenotetext{b}{Extrapolated as ${\cal A}\,(100\ \YeV) = {\cal A}_{\rm best} (E_{\rm best} / 100\ \YeV)^{-0.36}$, based on the growth of the cross section for neutrino-nucleon interaction \citep{Gandhi98}.  (In \citealt{Gayley09}, this is the decreasing neutrino interaction length.)}
\tablenotetext{c}{I multiplied the flux limits on $\tau$-flavor neutrinos and antineutrinos by 3 to get the all-flavor limits, under the assumption that the flavor ratio is near 1:1:1.}
\end{deluxetable*}

The highest sensitivity yet attained at YeV energies is by the Westerbork Synthesis Radio Telescope (WSRT) NuMoon experiment, which had an etendue of $\sim 0.01\ \oplus\,\sr$ at 100 YeV.  Since NuMoon had an exposure of only 46.7 hours \citep{Scholten09,Buitink10}, though, the limits are actually quite weak at YeV energies (as seen in Figure~\ref{fig:CurrentLimits}).  In the next 10 to 20 years, limits on YeV neutrinos will improve by a factor of 10 or more, when the Low Frequency Array (LOFAR; \citealt{Buitink13}) and Square Kilometer Array (SKA; \citealt{Bray14}) search the Moon for Cherenkov flashes.  The best etendue that ground-based lunar Cherenkov experiments can reach is limited by the surface area of the Moon's near side, $a = 0.044\ \oplus$.  Future experiments can detect radio pulses more deeply buried neutrino showers if they are more sensitive, but their etendue can be at most a factor of 10 greater than NuMoon.  The main improvement from LOFAR and SKA at Planck energies is simply that they will observe the Moon for longer periods (perhaps 1000 hours; \citealt{Bray14}).  

The method for deriving limits on the flux from point sources is similar.  The number of events expected in the experiment is then
\begin{equation}
\label{eqn:NEventsSources}
\mean{N_{\rm events}} = E \frac{dN}{dE} a_{\rm eff} t_{\rm exp},
\end{equation}
where $a_{\rm eff}$ takes into account the position of the source on the sky.  The sensitivity of neutrino experiments typically varies with direction, but order of magnitude estimates can be set by assuming a uniform sensitivity across the field of view, so that $a_{\rm eff} \approx {\cal A} / \Omega_{\rm eff}$.  Experiments observing the Moon at low radio frequencies ($\sim 100\ \MHz$) include WSRT and eventually LOFAR and SKA-Low.  It is thought that these experiments are sensitive to neutrinos hitting the Moon's surface coming from any inclination in the sky \citep{Scholten06}, in which case $\Omega_{\rm eff} \approx \pi\ \sr$.  Most of the other lunar Cherenkov experiments in Table~\ref{table:YeVExperiments} observe at GHz frequencies, in which case $\Omega_{\rm eff}$ is smaller.  This makes them more sensitive to point sources than their lower ${\cal A}$ implies, if their field of view includes the source.  LUNASKA, for example, is more sensitive to Centaurus A than NuMoon \citep{James11}.

\subsection{Direct experimental limits on a steady-state neutrino background}
\label{sec:SteadyBackground}
The expected number flux from K3 YeV neutrino sources is quite small, but not out of the question.  Suppose there is an accelerator lying a distance $D$ away that emits $10^{10}\ \Lsun$ entirely in 100 YeV neutrinos.  Then the expected neutrino flux is 
\begin{equation}
\phi \approx 8\ \Earthyr^{-1} \left(\frac{L}{10^{10}\ \Lsun}\right) \left(\frac{E}{100\ \YeV}\right)^{-1} \left(\frac{D}{100\ \Mpc}\right)^{-2}.  
\end{equation}
Such a source could be detectable in the intermediate future, if the entire Earth's surface is monitored for neutrinos, or if the Moon is observed for several years.  

However, current experimental limits on YeV neutrinos are much weaker than this flux.  From equation~\ref{eqn:NEventsSources}, NuMoon could have detected a $\ga 10^5\ \Lsun\,(E/100\ \YeV)$ neutrino source in the Galactic Center.  NuMoon could have missed a K3 neutrino source located as close as 2 Mpc.  LOFAR or SKA could improve the limits on Galactic Center neutrino sources to a few hundred $\Lsun$ at 100 YeV.  Individual K3 neutrino sources about 25 Mpc would be in reach.  Finally, an Earth-year of exposure is sufficient to detect individual K3 neutrino sources within about 300 Mpc at 100 YeV, or individual K2 neutrino sources within about 3 kpc.

\begin{figure*}
\centerline{\includegraphics[width=18cm]{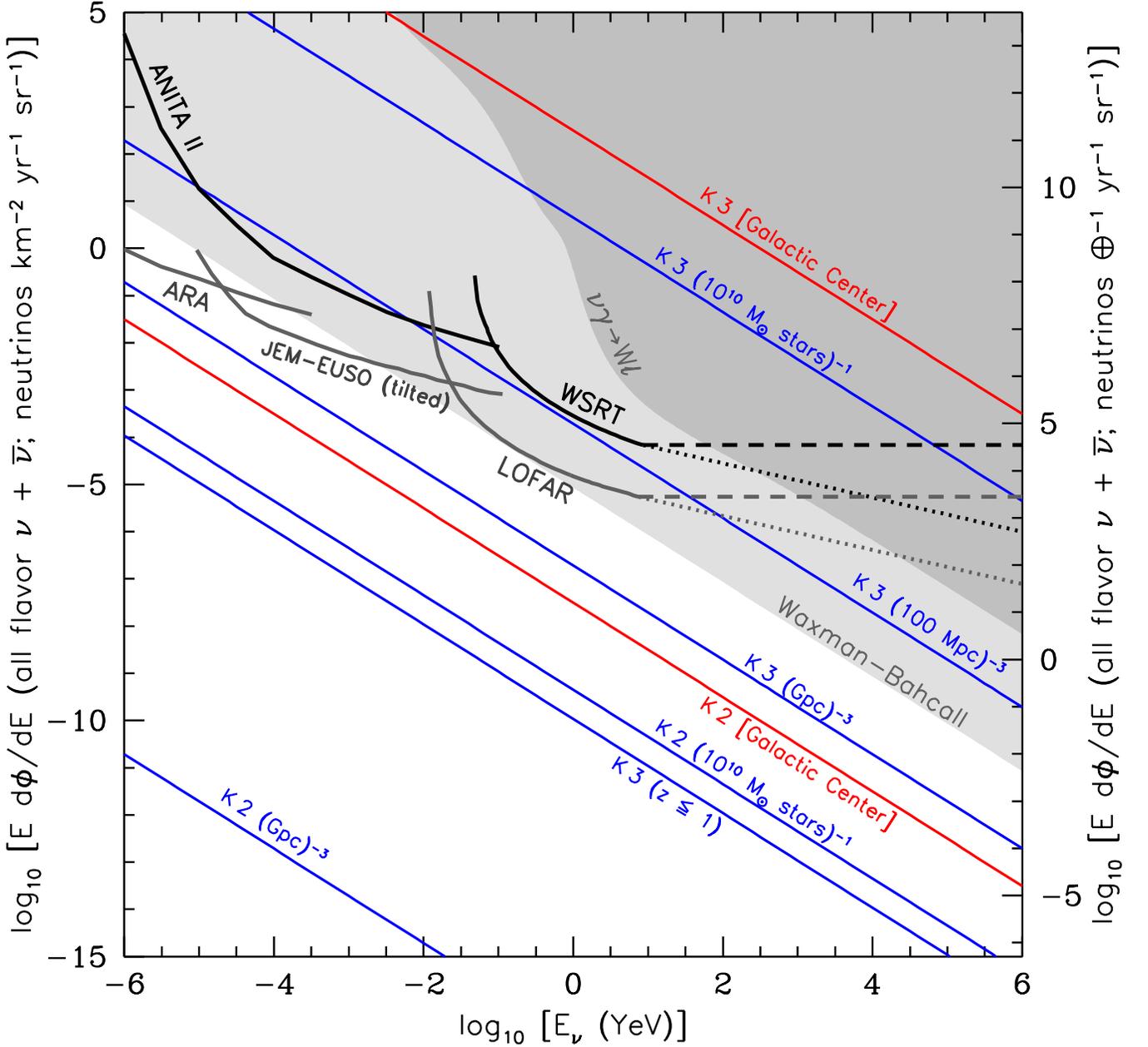}}
\figcaption{Current limits on the isotropic, steady-state background of YeV neutrinos at Earth.  The solid black lines are extant limits, while the grey solid lines are the projected limits from future experiments (references listed in Table~\ref{table:YeVExperiments}).  I extrapolate the limits from WSRT and LOFAR to XeV energies assuming a constant sensitivity (dashed lines) or as ${\cal A} \propto E^{-0.363}$ from the neutrino-nucleon cross section dependence (dotted lines).  Indirect limits on the cosmic background include constraints on emission of W-bursts (darker grey shading), and the Waxman-Bahcall limit (light grey shading; no evolution).  The indirect limits only apply to extragalactic backgrounds.  I also show the expected neutrino backgrounds from various extragalactic (blue) and Galactic (red) monoenergetic neutrino sources with various space densities and Kardashev ratings. The point source fluxes are converted into effective isotropic backgrounds using $\Omega_{\rm eff} = \pi\ \sr$.\label{fig:CurrentLimits}}
\end{figure*}

While individual accelerators are currently difficult to detect, we can also consider the integrated background from all accelerators in the Universe.  Although the flux of an individual source goes as $D^{-2}$, the number of sources goes as $D^2$, so most of the total flux comes from the sources distant from us.  To order of magnitude, the total neutrino background is
\begin{equation}
\label{eqn:nuBackground}
E^2 \frac{d\phi}{dE} = \frac{c n_{\rm source} L_{\rm source} t_{\rm source}}{4\pi},
\end{equation}
where $n_{\rm source}$ is the number density of sources, $L_{\rm source}$ is the luminosity of a typical source, and $t_{\rm source} \approx 10\ \Gyr$ is the time that sources have been present in the Universe.  I ignore evolution effects; star formation and AGN activity both were much more intense per comoving volume at high $z$, but presumably it takes some time for life to evolve, which could favor cosmic engineering being at low $z$.  

I show the expected background for monoenergetic neutrino sources with various Kardashev ratings and various densities as blue lines in Figure~\ref{fig:CurrentLimits}.  Current experiments imply that K3 neutrino sources are typically spaced $>100\ \Mpc$ apart, for all neutrino energies of $1\ \EeV \la E \la 3\ \YeV$.  Evidently, K3 YeV accelerators are either very rare, totally opaque, highly beamed, or intermittent.  In a typical 100 Mpc long cube, there are roughly $2 \times 10^{14}\ \Msun$ of stars, and $5000$ galaxies with stellar masses bigger than $10^{10}\ \Msun$ \citep{Baldry12} -- about 10\% of the number being studied by the \^G Infrared Search, which also has negative results \citep{Wright14-Results}.  This supports the idea that K3 entities are uncommon.

\subsection{Indirect limits on the isotropic steady-state neutrino background}
\label{sec:Indirect}
The weak limits on the highest energy neutrinos stem from the tiny expected number fluxes, but this does not mean the energy flux is small.  If a significant fraction of the energy in YeV--XeV neutrinos is converted into radiation of lower energy, it would be detectable with current instruments.  This indirectly limits the cosmic YeV--XeV neutrino background.

The ``natural'' strength of the neutrino background is the Waxman-Bahcall bound, $E^2 d\phi/dE = 2.7 \times 10^{-8}\ \GeV\ \cm^{-2}\ \sec^{-1}\ \sr^{-1}$ (\citealt{Waxman99}, after including all flavors and supposing there is no evolution; light grey shading in Figure~\ref{fig:CurrentLimits}).  It is essentially the flux in observed UHECRs.  The reasoning behind this bound is that if high energy neutrinos are accelerated in the same sources as UHECRs, the cosmic luminosity in neutrinos is at most the luminosity accelerated in UHECRs, assuming that the sources are transparent to UHECRs.  While its assumptions may not apply to artificial accelerators, it does set an upper limit on the natural background of neutrinos.  In fact, the background of PeV neutrinos observed by IceCube roughly lies on the Waxman-Bahcall bound \citep{Aartsen13,IceCube13}.  Experiments at other energies also aim to reach the bound within the next few years (Figure~\ref{fig:CurrentLimits}).  

The Waxman-Bahcall bound implies a small density of neutrino sources, no matter the neutrino energy (c.f. equation~\ref{eqn:nuBackground}):
\begin{equation}
n_{\rm source} \la 4 \times 10^{-8}\ \Mpc^{-3}\ \left(\frac{L_{\rm source}}{10^{10}\ \Lsun}\right)^{-1} \left(\frac{t_{\rm source}}{10\ \Gyr}\right)^{-1}.
\end{equation}
This corresponds to a spacing of $\ga 280\ \Mpc$ between K3 neutrino sources (a volume that includes $5 \times 10^{15}\ \Msun$ of stars, and about $10^5$ galaxies with $>10^{10}\ \Msun$ of stars), or $\ga 6\ \Mpc$ between K2.5 neutrino sources.  Shock CR acceleration produces particle spectra with a $dN/dE \propto E^{-2}$ or steeper spectra.  If this is how an artificial accelerator works, then it is necessarily brighter at the well-constrained PeV--ZeV energies than at YeV--XeV energies.  However, the Waxman-Bahcall bound does not apply if neutrino sources are monoenergetic.

A more robust limit can be set by the interaction of YeV--XeV neutrinos with the CMB.  Although the optical depth of the Universe to $\nu\gamma$ interactions is only $\sim 7 \times 10^{-4}$, it is not zero \citep{Seckel98}.  A small but potentially observable fraction of cosmic YeV--XeV neutrinos is converted into W bosons during their propagation to Earth.  The W bosons produce hadrons and leptons (a W-burst), with about half of the energy ultimately cascading down into GeV $\gamma$-rays. 

The observed GeV $\gamma$-ray background \citep{Abdo10,Ackermann15} thus limits the cosmic energy injection into neutrinos even at these energies \citep{Coppi97,Murase12}.  I calculate this limit on a neutrino background at YeV energies and plot it in Figure~\ref{fig:CurrentLimits}, including both the CMB and the extragalactic background light presented in \citep{Franceschini08}.  Numerically, the $\nu\gamma$ bound corresponds to an energy flux of $E^2 d\phi/dE \approx 2 \times 10^{-5}\ \GeV\ \cm^{-2}\ \sec^{-1}\ \sr^{-1}$ at all energies greater than YeV.  The resulting limit on the density of K3 neutrino sources is
\begin{equation}
n_{\rm source} \la 4 \times 10^{-5}\ \Mpc^{-3}\ \left(\frac{L_{\rm source}}{10^{10}\ \Lsun}\right)^{-1} \left(\frac{t_{\rm source}}{10\ \Gyr}\right)^{-1},
\end{equation}
or a spacing of $30\ \Mpc$.  While not as strong as the direct limits at energies below 1 YeV, it does confirm that steady, isotropically-radiating YeV--XeV neutrino sources are very rare, occurring in less than 1\% of massive galaxies (less than 1 per $\sim 6 \times 10^{12}\ \Msun$ of stars).  The limit could probably be improved by considering the interaction of YeV--XeV neutrinos with the relic cosmic neutrino background (the Z-burst mechanism; \citealt{Fargion99,Weiler99}), but this requires knowledge of the neutrino masses \citep{Roulet93}, so I do not include it.  

Note that these limits only apply to the cosmic averaged emission of neutrinos.  They do not exclude any given individual source close to or within the Galaxy.  Most of the constraining power comes from galaxies that are several Gpc away.

Individual YeV neutrino sources might be visible through the W-burst mechanism in GeV $\gamma$-rays, even if they have no intrinsic electromagnetic emission.  Note that more distant sources would have a brighter apparent $\gamma$-ray luminosity, because the optical depth for neutrinos along the sightline is larger \citep[c.f.,][]{Essey11}.  In practice, though, the expected flux even for a K3 YeV neutrino source is too small to see with current GeV telescopes like \emph{Fermi}-LAT.  

\subsection{Limits on YeV neutrino bursts}

Planck accelerators may not be steady sources, but could be brief transients as in the fireball scenario (section~\ref{sec:AcceleratorConditions}).  The assumption of steady sources applies if the lifetime of the sources $\Delta t$ is longer than the exposure time $t_{\rm exp}$ and if there is at least one source in the sky at any given time.  For comparison, typical neutrino experiments last for a few weeks to a few years (Table~\ref{table:YeVExperiments}).

There are three conditions for detecting the YeV neutrinos from a transient event.  The first is that a detection experiment must be active when the transient occurs.  The second is that the experiment must be pointed at the transient when it occurs.  The third is that the fluence of neutrinos must be big enough that the experiment is likely to detect at least one during the event.  Suppose an experiment is sensitive to bursts of a given luminosity out to a distance $D_{\rm max}$.  The number of transients it observes is
\begin{equation}
\label{eqn:NEventTransient}
N_{\rm events} = \Gamma_{\rm burst} t_{\rm exp} \Omega_{\rm eff} D_{\rm max}^3 / 3,
\end{equation}
where $\Gamma_{\rm burst}$ is the rate of the events per unit volume, $t_{\rm exp}$ is the exposure/active time of the experiment, and $\Omega_{\rm eff}$ is the effective field of view of an experiment.  This equation applies if the bursts are short, with $\Delta t \ll t_{\rm exp}$.  

Longer-lasting transients are more likely to be active when the experiment is on, but they are harder to detect since only a fraction of the neutrino fluence can be detected.  If $\Delta t \gg t_{\rm exp}$, the number of transients observed can be approximated by replacing $t_{\rm exp}$ with $\Delta t$ in equation~\ref{eqn:NEventTransient}, and adjusting $D_{\rm max}$ to reflect the new sensitivity.  The observed fluence is the total fluence multiplied by a factor $t_{\rm exp} / (\Delta t)$, however.  

The number of neutrinos generated can be related to the total integrated energy released during the event.  As shown in Table~\ref{table:AcceleratorPower}, a particle accelerator that can set useful constraints on Planck physics would generally need to convert $10^{-8}$--$100\ \Msun c^2$ of energy into YeV--XeV particles.  From energy conservation, $10^{40}$ 100 YeV neutrinos can be generated from each $\Msun c^2$ of energy in accelerated particles.  The fluence of an isotropic burst is 
\begin{equation}
\Phi = 120\ \oplus^{-1}\ \left(\frac{\Upsilon_{\nu}}{\Msun c^2}\right) \left(\frac{E}{100\ \YeV}\right)^{-1} \left(\frac{D}{\Gpc}\right)^{-2}.
\end{equation}

Current limits on neutrino bursts can therefore be calculated from the effective areas of neutrino experiments.  From Table~\ref{table:YeVExperiments}, UHE neutrino experiments range from those with huge effective areas but short exposure times (like NuMoon) to those with small effective areas but years of exposure (like RICE).  Bursts can be detected at distances of $D_{\rm max} \la [(\Upsilon_{\nu} a_{\rm eff}) / (4 \pi E_{\nu})]^{1/2}$.  In the sensitive/short exposure extreme, a $100\ \Msun c^2$ burst could have been detected by NuMoon out to $9\ \Gpc\,(E_{\nu} / 100\ \YeV)^{-1}$, ignoring cosmological effects.  A minimal $10^{-8}\ \Msun c^2$ burst from a neutrino-neutrino accelerator (Table~\ref{table:AcceleratorPower}) would have been detectable from a distance of $90\ \kpc\,(E_{\nu} / 100\ \YeV)^{-1}$, which is far enough to encompass the entire Galaxy and the Magellanic Clouds for $1\ \YeV \la E_{\nu} \la 300\ \YeV$.  RICE, the longest lasting YeV neutrino experiment, could only have detected a $100\ \Msun c^2$ burst out to $3\ \Mpc\,(E_{\nu} / 100\ \YeV)^{-1}$, only $10^{-10}$ of the volume probed by NuMoon.

In order to limit the rate at which short bursts occur per galaxy, we want to maximize the product of the probed volume ($\Omega_{\rm eff} D_{\rm max}^3$) and the exposure time.  A useful figure-of-merit is ${\cal M} = \Omega_{\rm eff} a_{\rm eff}^{3/2} t_{\rm exp}$.  I computed ${\cal M}$ for the experiments listed in Table~\ref{table:YeVExperiments}, using the extrapolated ${\cal A}$ at 100 YeV and with $\Omega_{\rm eff} = \pi\ \sr$.  NuMoon has the largest ${\cal M}$ by far, $5 \times 10^7\ \km^3\ \sr\ \yr$.  Of course, the range of NuMoon is so far that cosmological effects matter, so this cannot be taken as exact.  Yet, only FORTE comes anywhere near NuMoon, with ${\cal M} = 7 \times 10^6\ \km^3\ \sr\ \yr$.  ANITA, the third best, has ${\cal M} = 8 \times 10^4\ \km^3\ \sr\ \yr$; RICE has a mere ${\cal M} = 400\ \km^3\ \sr\ \yr$.  Larger effective areas are key.

LOFAR and SKA will have effective areas somewhat better than NuMoon and will run for a week to a month, with ${\cal M} \approx 10^9$ -- $10^{10}\ \km^3\ \sr\ \yr$.  For comparison, if a burst occurred in every $L_{\star}$ galaxy once in the past 10 Gyr, then there would be about 1 burst per year in the Hubble volume.  We are still far from ruling out a hadronic Planck accelerator being present in \emph{every} galaxy, if they are run in short bursts.

\subsection{Towards $1\ \oplus\ \yr$ exposures: Earth's oceans as an experimental volume}
\label{sec:Oceans}
Because YeV--XeV neutrinos are so rare, the Moon simply is not a big enough detector to observe distant colliders.  The only other planetary-sized body we can observe UHE neutrinos signals from the ground is the Earth itself.  About 70\% of Earth's surface is covered by the oceans, and their average depth is of order a few kilometers.  The columns of water in the oceans are big enough to stop YeV--XeV neutrinos, so the oceans are a good candidate experimental volume.

There are at least three potential signals from neutrino cascades in seawater.  The first is the radio Cherenkov pulses, as observed on the Moon.  But since seawater is highly conductive, radio emission is absorbed rapidly in the oceans, so these are undetectable.  The second is the optical Cherenkov flash, sought by the ANTARES neutrino telescope.  These experiments seek TeV--PeV neutrinos, though, and use small experimental volumes with closely spaced photodetectors \citep{Ageron11}.  They cannot be readily extended to YeV--XeV energies because the absorption length of optical light is a few hundred meters \citep{Aguilar05}.  Thus, the photodetectors cannot be more than $\sim 1\ \km$ apart.  For the cubic kilometer volumes of PeV neutrino telescopes, this is not a problem, but for all of the oceans, one would need billions of photodetectors.

The third option is to listen for the sound of a neutrino cascade \citep{Urick83}.  A 100 YeV neutrino dumps about 3 megajoules of energy (about 1 kg TNT equivalent) into a cubic meter of volume, assuming 20\% of its energy goes into a hadronic cascade \citep{Niess06}.  This heats the water, causing it to expand, and launches a pressure wave \citep{Askaryan77,Learned79,Lehtinen02}.  The characteristic frequencies are of order 10 kHz.  Acoustic experiments search for these sound pulses; the largest so far is SAUND \citep{Kurahashi10}.  The conversion of heat into sound is very inefficient \citep{Learned79,Price96}, and the sound is beamed \citep{Lehtinen02}, partly accounting for why the etendue of acoustic detectors are so far weak.  Present efforts are directed towards finding 100 EeV GZK neutrinos, whereas Planck neutrinos will be proportionally much louder, compensating for the low efficiency.  

Suppose that a fraction $\eta_{\rm sound}$ of the energy in the neutrino is emitted as a sound pulse with frequency $\nu_{\rm pulse}$.  The power of the pulse is approximately $\eta_{\rm sound} E \nu_{\rm pulse}$.  The sound is not emitted isotropically, but is concentrated in a disk (or ``pancake'') expanding perpendicularly from the neutrino cascade \citep{Lehtinen02,Vandenbroucke05}.  The anisotropy can be described by an angular distribution $d\bar{I}/d\Omega$, normalized so that $\int (d\bar{I}/d\Omega) d\Omega = 1$.  For hadronic showers, the emission disk has an angular thickness of order $\Delta \theta \approx 1^{\circ}$ \citep{Lehtinen02}, so $d\bar{I}/d\Omega \approx 100$ within the disk.  In addition, seawater absorbs sound with an energy attenuation length $D_{\rm abs}$.  Then the energy flux at a distance $D$ in the water and direction ${\bf k}$ is $I = [d\bar{I}({\bf k})/d\Omega] \eta_{\rm sound} E \nu_{\rm pulse} \exp(-D/D_{\rm abs}) / D^2$.  

The pressure difference is related to the energy flux as $I = p^2 / (\rho c_{\rm sound})$, where $\rho c_{\rm sound} = 1.5 \times 10^5\ \gcm2\ \sec^{-1}$ in water \citep{Urick83}.  We have
\begin{equation}
\Delta P = \left[\frac{d\bar{I}({\bf k})}{d\Omega} \frac{\eta_{\rm sound} E \mean{\nu_{\rm pulse}} \rho c_{\rm sound}}{4 \pi D^2} \exp\left(-\frac{D}{D_{\rm abs}}\right)\right]^{1/2}.
\end{equation}
For a 100 YeV neutrino detected at a distance, the pressure wave has an amplitude of roughly
\begin{align}
\label{eqn:pAcoustic}
\nonumber \Delta P  & = 1.4\ \Pa\ \left(\frac{D}{100\ \km}\right)^{-1} \exp\left(-\frac{D}{2 D_{\rm abs}}\right)\\
                    & \times \left[\left(\frac{d\bar{I}({\bf k})/d\Omega}{100}\right)\left(\frac{\eta_{\rm sound}}{10^{-8}}\right)\left(\frac{E}{100\ \YeV}\right) \left(\frac{\mean{\nu_{\rm pulse}}}{10\ \kHz}\right)\right]^{1/2}.
\end{align}
I estimate $\eta_{\rm sound}$ from the pressure profiles given in \citet{Lehtinen02}.  An acoustic array operating at these frequencies has an expected pressure sensitivity of $1\ \mPa$ \citep{Lehtinen02} to $0.1\ \Pa$ \citep{Kurahashi10}.

The first limiting factor for an acoustic array is sound absorption in seawater, since the energy attenuation lengths are a few kilometers or less above 10 kHz  \citep{Fisher77,Francois82a,Francois82b,Ainslie98}.  Under some standard assumptions about ocean temperature ($T = 2^{\circ} {\rm C}$), salinity ($S = 0.035$), acidity (${\rm pH} = 8$), and depth ($z = 5\ \km$), the ocean attenuates sound at a rate
\begin{equation}
\left(\frac{\alpha_{\rm sound}}{\dB/\km}\right) \approx \left\{ \begin{array}{lr} 0.13 \nu_{\rm kHz}^2 & (\nu_{\rm kHz}^2 \la 0.84)\\\\
                                                       0.089 + 0.0053 \nu_{\rm kHz}^2 & (0.84 \la \nu_{\rm kHz}^2 \la 45)\\\\
                                                       11 + 3.4 \times 10^{-4}\ \nu_{\rm kHz}^2 & (45 \la \nu_{\rm kHz}^2) \end{array} \right.,
\end{equation}
according to \citet{Ainslie98}, where $\nu_{\rm kHz}$ is the sound frequency in kHz.  This can be translated into an energy attenuation length: $D_{\rm abs} = 10\ \dB / (\alpha_{\rm sound} \ln 10) = 4.3\ \km\ [\alpha_{\rm sound} / (\dB\ \km^{-1})]^{-1}$.  At 10 kHz, the attenuation length is about 7 kilometers.  Thus, the exponential factor in equation~\ref{eqn:pAcoustic} is $0.03$ for $D = 50\ \km$, $8 \times 10^{-4}$ for $D = 100\ \km$, and $2 \times 10^{-5}$ for $D = 150\ \km$.  Despite the enormous losses, the pulse from a 100 YeV neutrino is so loud that it could be detected $D_{\rm range} \approx 45\ \km$ away by an array sensitive to 0.01 Pa sounds. 

The other limitation is the narrowness of the acoustic disk.  If a neutrino comes from the zenith, the acoustic disk expands outwards into the sea without interacting with the seafloor and sea surface.  But if a neutrino comes from the horizon, most of the sound is directed towards the seafloor and sea surface.  A neutrino must have a zenith angle of $\la z / D_{\rm range}$ to be regarded as horizontal, where $z$ is the ocean depth.  For $z \approx 5\ \km$, this limits the zenith angle to about $6^{\circ}$, or only 0.3\% of the celestial sphere.  A completely vertical neutrino illuminates only $\Delta \theta$ in azimuth out to large distances.  

If we were only interested in detecting zenith neutrinos, then we could settle for a few detectors per $D_{\rm range}^2$.  To cover the Earth's oceans would require $A_{\oplus} / D_{\rm range}^2 \approx 2 \times 10^5$ detectors.  For comparison, the SAUND II array used only 49 hydrophones \citep{Kurahashi10}.  Furthermore, these detectors would actually be strings of hydrophones, to ensure that one intersects the acoustic disk \citep{Vandenbroucke05}.  But since the distance between the strings is big, the hydrophones can be spaced about one per kilometer on the strings.

The effective etendue of an ocean array that detected only zenith neutrinos would actually be less than NuMoon, although a hydrophone array that covered the Earth could observe in all directions and have a longer livetime.  To effectively detect horizon neutrinos on-disk, the number of detector strings must be multiplied by $2 \pi / \Delta \theta \approx 400$, for a grand total of $10^8$ strings.  

A much more promising route is to consider the off-disk acoustic emission from a shower.  Because of the exponential attenuation factor, the  range only depends logarithmically on $d\bar{I}/d\Omega$.  According to \citet{Lehtinen02}, the sound fluence of a hadronic shower $5^{\circ}$ away from the disk is about 0.1\% of maximum.  Unfortunately, the sound emission at larger angles is not given.  As a fiducial estimate, suppose $d\bar{I}/d\Omega = 10^{-3}$.  The pressure differential then remains above 1 Pa out to $\sim 15\ \km$ for a 100 YeV neutrino (0.01 Pa sensitivity), increasing the number of detectors by a factor of $\sim 10$ only.  For a Planck neutrino, the range increases to $\sim 30\ \km$ (0.01 Pa sensitivity).  Furthermore, the detectors would not have to be strings, since they would be listening for the isotropic sound emission.

It may be worthwhile to study the low frequency component of these pulses with care.  Sounds below a kHz can propagate thousands of kilometers, so a very sparse array might be able to detect the cascades of the highest energy neutrinos.

\subsection{The far future of YeV neutrino detection: how big can we go?}
\label{sec:BigExperiments}
It would take exposures of Earth-centuries or millennia to rule out K3 YeV accelerators anywhere in the Universe, much less those with the minimal luminosities listed in Table~\ref{table:AcceleratorPower}.  But must we limit ourselves to Earth, if we are going to extrapolate that far into the future?

\subsubsection{Jupiter (a bad idea)}
Jupiter has a surface area that is $\sim 100$ times bigger than that of Earth.  Could we deploy acoustic detectors throughout Jupiter's atmosphere to listen for YeV neutrinos?  I will be fanciful for a moment about our capabilities this far in the future and ignore the logistics.  Maybe the detectors can be quickly deployed by launching them on self-replicating machines, creating an ecology of ``floaters'' once speculated to live in Jupiter \citep{Sagan76}.

The basic challenge facing any such attempt at instrumentation is that the density of an atmosphere is low while the column density to stop a YeV neutrino is relatively high (equation~\ref{eqn:UHENuColumn}).  The outer thousand kilometers of Jupiter's atmosphere is approximately adiabatic with $P = P_0 (\rho / \rho_0)^{5/3}$ \citep{Seiff96}.  From the equation of hydrostatic stability, $dP/dz = \rho g$, the density in Jupiter's atmosphere at depth $z$ is
\begin{equation}
\rho = \left[\frac{2}{5} \frac{g\rho_0}{P_0} z\right]^{3/2} \rho_0.
\end{equation}
The column of the atmosphere above depth $z$ is given by $d\Sigma/dz = \rho$, so that $\Sigma = (2/5) \rho z$.  

Define as a reference ``surface'' the 1 bar depth of Jupiter's atmosphere, where $P_0 = 10^6\ \dyncm2$, $\rho_0 \approx 0.0002\ \rhoUnits$, and the temperature is $T_0 \approx 170\ \Kelv$ \citep{Seiff96}.  The gravitational acceleration in the outer parts of Jupiter is $g = 2500\ \cms2$.  Neutrinos interact at a typical depth of 
\begin{align}
\nonumber z_{\nu} & = \frac{5}{2 \rho_0} \left(\frac{P_0}{g}\right)^{3/5} \Sigma_{\nu}^{2/5} = 1100\ \km\ \left(\frac{\Sigma_{\nu}}{10^6\ \gcm2}\right)^{2/5},
\end{align}
where the density is $\rho_{\nu} = 0.022\ \rhoUnits\,(\Sigma_{\nu} / 10^6\ \gcm2)^{3/5}$ and the pressure is $P_{\nu} = 2500\ \atm \,(\Sigma_{\nu} / 10^6\ \gcm2)$.  The temperatures at these depths are 
\begin{equation}
T_{\nu} = T_0 \left(\frac{P_{\nu}}{P_0}\right)^{2/5} = 3900\ \Kelv\ \left(\frac{\Sigma_{\nu}}{10^6\ \gcm2}\right)^{2/5}.
\end{equation}
Jupiter's atmosphere may be nonadiabatic at depths greater than $1000\ \atm$, and the temperature could be somewhat lower \citep{Guillot94}, but in any case $z_{\nu}$ is extremely deep.

These physical conditions present enormous challenges to detecting neutrinos.  The pressure is a few times that in Earth's deepest ocean trenches, but the temperature is far beyond even the surface of Venus.  No known electronics can withstand $\gg 1000\ \Kelv$ temperatures for any significant time; indeed, most solid materials melt at these depths.  Even if the material engineering challenges could be overcome, the high temperatures lead to a huge thermal background of noise that would cover up neutrino signals.  Similar considerations apply for the other gas giants, not to mention the Sun.

Gaseous atmospheres are a poor choice for a YeV neutrino detector target because they are too compressible.  The pressure in an incompressible liquid ocean made of water builds up rapidly, as its density does not change with depth.  The pressure and density of a liquid is basically independent of its temperature.  But a gas can support the increased pressure load by becoming hotter without becoming much denser.  Because of the much lower gas densities, neutrino absorption occurs very deep in the atmosphere, where it is extremely hot.

\subsubsection{A million Kuiper Belt Objects (a much better idea)}
The minor bodies of the Solar System collectively have more surface area than the major bodies.  Moreover, they are solid, so that $10^6\ \gcm2$ columns are achieved in all minor bodies with radii $R \ga 10\ \km$.  Detecting neutrinos in these icy objects does not even require landing on them.  An orbiting satellite can detect radio Cherenkov pulses from neutrino showers within them, as proposed for Europa \citep{Shoji11,Miller12} and Enceladus \citep{Shoji12}.

In terms of mass, the Kuiper Belt is the largest reservoir of minor bodies within 100 AU.  Collectively, they have a large surface area.  The radius distribution of Kuiper Belt objects (KBOs) is
\begin{equation}
\frac{dN_{\rm KBO}}{dR} \approx 710\ \km^{-1} \left(\frac{R}{45\ \km}\right)^{-4},
\end{equation}
for $0.25\ \km \le R \le 45\ \km$, as determined by stellar occultations \citep{Schlichting12}.  This integrates to a collective surface area of 
\begin{equation}
a_{\rm eff} = \int_{10\ \km}^{45\ \km} \frac{dN_{\rm KBO}}{dR} 4 \pi R^2 dR = 2.8 \times 10^9\ \km^2 = 5.6\ \oplus,
\end{equation}
for KBOs with $10\ \km \le R \le 45\ \km$.  Smaller objects have an exposure penalty, since the probability for neutrino interaction goes as $R$ when the column depth is less than $\Sigma_{\nu}$: $a_{\rm eff}^{\rm small} \approx \int 4 \pi R^2 (R / 10\ \km)\,(dN_{\rm KBO}/dR)\,dR \propto \ln R$.  I find that there is roughly $16\ \oplus$ in surface area on KBOs with $1\ \km \le R \le 10\ \km$, and another $16\ \oplus$ on KBOs with $0.1\ \km \le R \le 10\ \km$.  Far greater populations may exist in the Oort cloud, although they are much harder to reach.

A future civilization may instrument the larger KBOs to detect YeV neutrinos, attaining an exposure rate about $\sim 10$ times greater than that from the Earth alone.  If this becomes practical within a few centuries, the total integrated exposure to YeV neutrinos could be increased rapidly with a few decades of observation time.  The main practical challenge is the sheer number of objects that would need to be observed.  There are roughly
\begin{equation}
N_{\rm KBO} = \int_{10\ \km}^{45\ \km} \frac{dN_{\rm KBO}}{dR} dR \approx 1 \times 10^6
\end{equation}
KBOs with radii between 10 and 45 km, and about $10^9$ with radii between 1 km and 10 km.  Merely finding all of these objects, much less visiting them all, is a formidable challenge.  Unlike the case of Jupiter, no miraculous advances in material science are needed, though.

\section{Discussion}
\subsection{Summary}
Do some aliens leave their home planets and go on to cosmic engineering, do they all die before they can try, or do they all decide to stay at home?  The Fermi Paradox implies the first possibility is always wrong if ETs exist in any numbers.  Otherwise it is evidence that ETs are very rare.  One issue with the argument is that most of the proposed reasons for cosmic engineering involve either contacting young, human-like societies across the Universe, or consuming cosmic amounts of energy.  Remaking stars and galaxies to get our attention can seem a bit extravagant to get our attention, and consuming everything within reach is a questionable project if ETs care about sustainability and avoiding aggression with other ETs.  But I suggest there's at least one more motivation, curiosity: \emph{some scientific problems can only be answered with cosmic engineering}.

The Nightmare of particle physics is the dream of astronomers searching for ETs.  Planck accelerators must be bigger than the Sun, for realistic electromagnetic field strengths (Section~\ref{sec:BigAccelerators}).  The power used by a Planck accelerator is cosmic, because the natural cross section of Planck physics is tiny, with only one Planck event occurring per $\sim 10^{41}$ collisions (Section~\ref{sec:BrightAccelerators}).  The luminosities necessary for a hadronic collider in particular requires a K3 entity, if the experiment is to be completed within a million years (Table~\ref{table:AcceleratorPower}).  Whether or not they feel like communicating, we can still look for these kinds of energy usages.

The conditions in the accelerator region may be very hostile if it is a dirty fireball with a compactness problem, thermalizing all of the high energy particles.  Mesons and muons can be cooled before they decay into neutrinos, although we can still hope to see prompt neutrinos from kaons, charm, and beauty.  Alternatively, YeV--XeV radiation generated by the collisions could escape if the collider is clean or highly relativistic (Section~\ref{sec:AcceleratorConditions}).  Cosmic ray nucleons, photons, and $e^{\pm}$ can be detected from tens of Mpc away if they escape, and neutrinos are not stopped at all (Section~\ref{sec:Propagation}).  If nothing else, we can find Planck accelerators by their thermal emission, although it may emerge at much higher energies than the mid-infrared wavelengths of Dyson spheres (Section~\ref{sec:HiddenAccelerators}).

Our current knowledge of the high energy neutrino backgrounds implies that K3 YeV neutrino sources are typically at least 30 Mpc apart, if they are steady and radiate isotropically (Sections~\ref{sec:SteadyBackground} and~\ref{sec:Indirect}).  Individual K3 accelerators can only be detected out to about 2 Mpc.  In addition, there is no K\,2.5 or brighter 100 YeV neutrino source within about 10 kpc, implying there are no active Planck accelerators in the Galaxy (Section~\ref{sec:SteadyBackground}).  LOFAR searches for radio Cherenkov pulses on the Moon can improve YeV neutrino exposures by a factor of ten to a hundred, mainly because of a longer exposure time (Section~\ref{sec:CurrentExperiments}).  If we could place hydrophone arrays throughout the Earth's oceans and listen for cascade acoustic signals for a year, we could improve the bounds by another factor of a thousand.  There are many practical difficulties, though, particularly the $10^5$ hydrophones required and the beaming of the acoustic signals (Section~\ref{sec:Oceans}).  Exposures much beyond an Earth-year might be achieved in the far future by searching with satellites for radio Cherenkov pulses in all $10^6$ Kuiper Belt objects wider than 20 km (Section~\ref{sec:BigExperiments}).  

\subsection{Where is their nonthermal emission?}

The Universe emits very little nonthermal emission -- the lack of observed YeV neutrinos is just one symptom of this fact.  If the Universe were filled with diverse cosmic engineers with varying motives, one might figure that there would be artificial emission across all energies and messengers.  Some of their projects might generate radio broadcasts; other might generate $\gamma$-ray emission; others still neutrinos or gravitational waves -- a sort of Copernican principle for messengers.  

But almost all of the luminosity of the Universe is thermal, with $u_{\rm EBL} \approx 3 \times 10^{-14}\ \erg\ \cm^{-3}$ in the infrared to ultraviolet bands \citep{Franceschini08}, and a similar density in thermal MeV neutrinos emitted during core-collapse supernovae \citep{Beacom10}.  The characteristic energy density of UHECRs and neutrinos, the Waxman-Bahcall limit, is a \emph{million} times smaller \citep{Waxman99}.  Likewise, the GeV $\gamma$-ray background has a density that is $\sim 10^{-5}$ that of the EBL \citep{Ackermann15}.  While the intensity of the extragalactic GHz radio background is disputed, it is at most $4 \times 10^{-6}$ the density of the EBL \citep{Fixsen11} and perhaps a factor of 10 below that.  At lower frequencies, the constraints just become even tighter, with $\nu u_{\nu} \propto \nu^{-0.3}$ \citep{Fixsen11}.  

While it is true that most of the power used in cosmic engineering must be thermal emission, if even one in a thousand galaxies had a K3 entity that emitted 1\% of its power in one of these bands, they would overproduce the observed background radiation.  The problem is especially severe at radio wavelengths, where we can now observe star-forming galaxies out to high redshift.  Yet most star-forming galaxies appear to lie on the far-infrared--radio correlation, indicating natural synchrotron and free-free radio emission \citep{Yun01,Mao11}.  K3 radio emitters would make these galaxies \emph{brighter} at radio wavelengths and easier to see, but we have not found any artificial extragalactic radio (or $\gamma$-ray or neutrino) sources.  Cosmic engineering's effect on the Universe appears to be utterly minuscule, and the Fermi Paradox remains mystifying.

\subsection{Neutrino SETI in context}

Although searches for UHECRs and neutrinos of all energies from ETs have been proposed in the past \citep{Subotowicz79,Learned94,Swain06,Silagadze08,Learned09}, there has not been a dedicated effort to do them, unlike in radio and optical.  There are crucial differences between neutrino and radio/optical astronomy that are relevant.  Radio and optical telescopes use \emph{focusing optics}.  Their fields of view are small, so they must be aimed, except for something like the Fast Fourier Transform Telescope \citep{Tegmark09}.  In addition, individual photons are easy to detect.  The backgrounds from natural sources and the Galaxy mean that one must carefully search for desired signals.  A radio/optical telescope cannot be simply turned on and left to search the entire sky for any signal from all sources.  Observations need specific goals in these wavebands.  If SETI is not an explicit goal of the observers, one must settle for a ``commensal'' program, where one searches for artificial signals within the field of view of the primary targets, hoping that ETs happen to lie in those directions \citep{Bowyer83,Siemion14}.  

In contrast, neutrino telescopes cannot be aimed.  While the geometry of the Earth or the Moon can target some of the sky, their fields of view are nonetheless vast.  We cannot focus high energy neutrinos; nor can we move cubic kilometers of ice or water or regolith, much less entire planets.  Thus, there is no real way to be specific about the targets of a neutrino telescope.  The main challenge instead is simply detecting individual neutrinos, just because they are so rare, and figuring out where they come from.  Any experiment looking for high energy neutrinos is a SETI experiment, since every observing program is commensal with every other for high energy neutrinos.  As experiments across the neutrino spectrum, from TeV to YeV, reach the astrophysically motivated Waxman-Bahcall bound, neutrino SETI automatically begins.

To be clear, Planck scale SETI is not a replacement for more traditional radio and optical SETI.  Rather the two are complimentary \citep{Wright14-SF}.  We don't actually know cosmic engineers can exist, but we do know of one radio- and optical-broadcasting society, our own.  Conversely, even if cosmic engineers are rare, they can be seen across the Universe.  Furthermore, radio wavelengths may be particularly useful for searching for K3 entities, just because galaxies are typically so faint in radio and the sensitivity of current radio instruments is far greater than in $\gamma$-rays or neutrinos.

Detection of YeV radiation of any kind would imply a completely new class of astrophysical objects, either an unprecedented particle accelerator \citep{Thompson11} or something even weirder like a topological defect.  Whether or not cosmic engineers exist, the search for Planck scale particles in the Universe will be exciting.

\acknowledgments
I acknowledge support from the Institute for Advanced Study.

\end{document}